\documentclass[11pt]{article}
  
\usepackage{amsmath,amssymb}

\usepackage{epsfig}  
\usepackage{graphicx}               
\usepackage{url}
\usepackage{cite}
\usepackage{hyperref}
\usepackage{bm}

\usepackage{pstricks}
\usepackage{color}
\usepackage{multirow}
\usepackage{slashed}
 
\newcommand{\nc}{\newcommand}  

\nc{\beq}{\begin{equation}}  
\nc{\eeq}{\end{equation}}  
\nc{\beqa}{\begin{eqnarray}}  
\nc{\eeqa}{\end{eqnarray}}  
\nc{\bit}{\begin{itemize}}  
\nc{\eit}{\end{itemize}}

\title{  
\vspace*{-2.3cm}
\begin{flushright}
\normalsize{
  }
\end{flushright}
\vspace{1.5cm}
\Large  
\textbf{Superbumps
}\vspace*{1.0cm}   
}

\author{Yang Bai and Joshua Berger
\vspace{5mm}
\\
\normalsize\emph{Department of Physics, University of Wisconsin-Madison, Madison, WI 53706, USA}  
}
\date{}

\begin{document} 

\setcounter{page}{0}  
\maketitle  

\vspace*{1cm}  
\begin{abstract} 
For a wide range of supersymmetric models, there is a chiral superfield whose scalar and pseudo-scalar have approximately degenerate masses and couplings to Standard Model particles. At colliders, they may show up as ``superbumps": a pair of resonances with similar masses and production cross-sections. Observing the superbumps may provide evidence of supersymmetry even without seeing superpartners with a different spin. We present two models which realize the superbump scenario. The first one contains an elementary superfield, 24, under $SU(5)_{\rm GUT}$, while the second one is based on the supersymmetric QCD model with $N_f = N_c +1$ and identifying $SU(N_f=5)$ as $SU(5)_{\rm GUT}$. Both models have rich phenomenology including nearly mass-degenerate scalar and pseudo-scalar color octets that appear as three body resonances of two photons and one gluon. We also show that the recent 750 GeV diphoton excess at the LHC could be the first hint of a superbump signature. 
\end{abstract}  
  
\thispagestyle{empty}  
\newpage  
  
\setcounter{page}{1}

\baselineskip18pt   

\vspace{-2cm}

\section{Introduction}
\label{sec:intro}
Supersymmetry was discovered more than forty years ago from seeking a beautiful theory~\cite{Ramond:1971gb,Neveu:1971iv,Golfand:1971iw,Wess:1974tw} and from explaining light neutrinos~\cite{Volkov:1972jx}. Later, it was realized that the supersymmetry can be used to solve the ``hierarchy problem'' of the Standard Model (SM) Higgs boson.  The minimal solution, known as the Minimal Supersymmetric Standard Model (MSSM)~\cite{Martin:1997ns,Chung:2003fi}, predicts superpartners with masses at the TeV scale.  Furthermore, the lightest R-parity odd superparticle can be a Weakly Interacting Massive Particle (WIMP) and is a viable dark matter candidate~\cite{Feng:2010gw}. However, current searches at the Large Hadron Collider (LHC) for the superpartners have not shown any signs of existence of SM particle superpartners below around 1 TeV in mass~\cite{Aad:2016jxj}. Similarly strong constraints have been obtained from dark matter direct detection experiments, where the preferred SUSY dark matter parameter space has become increasingly excluded~\cite{Akerib:2015rjg,Cahill-Rowley:2014boa}. The current status of searches for supersymmetry motivates us to explore other ways to find signals of supersymmetry. 

Specifically, one could ask whether one can find the evidence of supersymmetry without discovering a superparticle. In this paper, we want to point out this possibility and explore the models and their particle mass spectrum for this to happen. The starting point of our observation is the following CP-conserving interaction of a complex scalar field, $X \equiv (X_R + i \, X_I)/\sqrt{2}$, with SM gauge boson field tensor
\beqa
 \frac{1}{M_{\rm UV}}\,\left( X_R\,W_{\mu\nu} W^{\mu\nu} \, + \, X_I\,W_{\mu\nu} \widetilde{W}^{\mu\nu} \right)\, + \,\frac{1}{2}\,M_X^2\, (X_R^2 \, + X_I^2) \,,
 \label{eq:relation}
\eeqa
where we have neglected gauge indices. $W_{\mu\nu}$ are SM gauge boson field tensors and $\widetilde{W}^{\mu\nu} = \frac{1}{2}\epsilon^{\mu\nu\alpha\beta}V_{\alpha\beta}$; the mass scale $M_{\rm UV}$ is related to some ultra-violet physics generating the dimension five operators. Notice that we have chosen the same mass and the same interaction strength for the scalar and pseudo-scalar states. There should be a symmetry that enforces this relation and protects it from radiative corrections. In the absence of additional Lagrangian terms, one could define a duality transformation associated with the source-free Maxwell equations, under which $(W_{\mu\nu} + i \widetilde{W}_{\mu\nu}) \rightarrow e^{i\theta}(W_{\mu\nu} + i \widetilde{W}_{\mu\nu})$, in addition to a simultaneous rotation $X \rightarrow e^{-2 i\theta} X$.  The interactions in Eq.~\eqref{eq:relation} are invariant under this transformation. The introduction of particles charged under the SM gauge symmetries breaks this enhanced duality symmetry.  Additional symmetry-breaking operators should therefore appear in the Lagrangian, following 't Hooft's Naturalness argument~\cite{naturalness}. For instance, the mass operators of the form $\mu^2\,(X^2 + X^{\dagger\, 2})$ explicitly break the above symmetry and split the scalar and pseudo-scalar particle masses. 

The story is different in a supersymmetric world. The perturbative non-renormalization theorem for the superpotential can protect relations in Eq.~(\ref{eq:relation}) from large perturbative radiative corrections~\cite{Grisaru:1979wc,Seiberg:1993vc}. The simplest way to see this is to treat the complex scalar $X$ as a part of a chiral superfield ${\bf X}$ and the gauge field as a part of chiral field strength superfield ${\bf W}_\alpha = \lambda_\alpha + \theta_\alpha D + \frac{i}{2}(\sigma^\mu \overline{\sigma}^\nu \theta)_\alpha W_{\mu\nu}   + i \theta\theta(\sigma^\mu \partial_\mu \lambda^\dagger)_\alpha$ in superspace notation. The interactions in Eq.~(\ref{eq:relation}) can then be matched to its supersymmetric version as $\int d^2\theta \,\left[- {\bf X}{\bf W}_\alpha {\bf W}^\alpha/(\sqrt{2} M_{\rm UV}) + M_X\, {\bf X}^2/2 + \mbox{h.c.}\right]$.  Holomorphy of the superpotential prevents additional terms that could split either the masses or interactions of the bosonic components of ${\bf X}$ in the absence of SUSY breaking effects. The supersymmetric interaction term introduces additional interactions for the fermionic superpartners of $X$ and various gauginos. It is not surprising that discovering the superpartners of $X$ and the gauge bosons can lead to a confirmation of the existence of supersymmetry. On the other hand, if we can discover the two scalar fields and test their mass and coupling relations, we may also infer the existence of supersymmetry without actually seeing the superparticles. 

\begin{figure}[th!b]
  \centering
\includegraphics[width=0.6\textwidth]{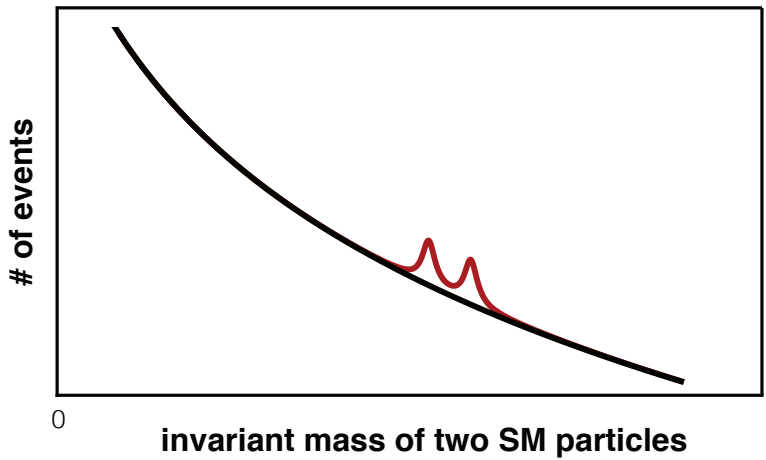}
  \caption{Schematic plot for a ``superbump" signature on top of the SM background.}
  \label{fig:schematic}
\end{figure}

Of course, supersymmetry is ultimately broken in our world. Supersymmetry breaking effects can feed into the $X$ particle sector. Assuming the SUSY-breaking effects are communicated to the $X$ sector by some heavy mediators, the modifications on the relations in Eq.~(\ref{eq:relation}) are suppressed and leave the mass and coupling relations approximately intact.  The two approximately-degenerate resonances could generically appear as a single broad resonance or two nearby separate narrow resonances, depending on the experimental resolution for the particles in their decays. In the latter case, we can have the schematic signal plus background event distributions in Fig.~\ref{fig:schematic}. 
In general, the two SM particles which make up the superbumps may not be two SM gauge bosons, although we present the interaction example in Eq.~(\ref{eq:relation}). There are also examples of superbumps reconstructed from two SM fermions, which we will come back to at the end of this paper.  Even in the Higgs field sector of the MSSM, one could identify a superbump scenario. In the decoupling limit with $M_{A^0} \gg M_Z$, the heavy Higgs boson, $H^0$, and the pseudo-scalar boson, $A^0$, become nearly degenerate in masses and couplings to SM fermions, with the relative splitting suppressed by $M_Z^2/M^2_{A^0}$. For a wide range of values of $\tan{\beta}$, one could search for superbumps from invariant mass distributions of $\tau^+ \tau^-$.  $H^0$ and $A^0$ also couple to two gluons or photons following the interactions in Eq.~(\ref{eq:relation}) at leading order in $M_Z^2/M^2_{A^0}$.  Although this serves as an existing example of a superbump signature in terms of two SM gauge bosons, the branching ratios are too small to be feasible at the LHC. Therefore, in this paper we present two more superbump models that can be realistically seen at the LHC and study correlated signatures in other channels.

For the first example, we take $X$ to be an elementary particle and take it to be one component of a chiral superfield that transforms as a {\bf 24} under the $SU(5)_{\rm GUT}$.  Superpotential interactions among the components of the {\bf 24} are sufficient to generate interactions of the SM-singlet component of the form in Eq.~(\ref{eq:relation}).  The soft masses in the MSSM, particularly the gluino mass, can radiatively split the masses and couplings of the scalar and pseudo-scalar components of $X$. For the second example, we take $X$ to be a meson of an $SU(N_c)$ supersymmetric QCD (SQCD) sector. While we have found that both the cases of $N_f= N_c +1$ and $N_f = N_c$ can realize a superbump signature, we choose $N_f=N_c + 1$ as a working example. For both the elementary and composite examples, we study the collider signatures associated with the SM charged particles as well as the neutral ones. The neutral states are of particular interest, as they can be seen as superbumps in the diphoton channel. Although our motivation is the bottom-up and signature-driven, we also note that the fermionic partner of $X$ could be the lightest supersymmetric particle (LSP) and serves as an alternative WIMP candidate in supersymmetric models. 

Our paper is organized as follows. In Section~\ref{sec:24}, we first introduce the simple {\small 24}MSSM model and study the MSSM-mediated SUSY-breaking effects that split the scalar and pseudo-scalar masses. We then present the composite model based on the SQCD with $N_f = N_c +1$ in Section~\ref{sec:model}, where two kinds of SUSY-breaking effects have been considered. In Section~\ref{sec:pheno}, we study the phenomenology of superbumps with high or low confinement scales including the potential 750 GeV diphoton superbumps. We discuss the additional signatures associated with the fermion degrees of freedom and other types of superbumps in Section~\ref{sec:conclusion}. In the appendix, we discuss the anomaly induced superpartner couplings, supersymmetric higher order K\"ahler potential and SUSY breaking K\"ahler corrections.

\section{A Simple {\small 24}MSSM model}
\label{sec:24}
As a first model, we consider the MSSM with one additional
field.  We take the new field $\mathbf{\Phi}$ to be a {\bf 24} under $SU(5)_{\rm
  GUT}$, which decomposes into SM representations $[SU(3)_C,SU(2)_W]_{U(1)_Y}$ as
\beq
\mathbf{24} = (\mathbf{8},\mathbf{1})_0 \oplus
(\mathbf{3},\mathbf{2})_{-5/6}  \oplus
(\overline{\mathbf{3}},\mathbf{2})_{5/6} \oplus
(\mathbf{1},\mathbf{3})_0 \oplus (\mathbf{1},\mathbf{1})_0 \,.
\eeq
At renormalizable level, the gauge-invariant terms of $\bf \Phi$ in the superpotential are
\beq
W = \frac{1}{2} M \mathbf{\Phi}^A \mathbf{\Phi}^A + \frac{\lambda}{4}\,d^{ABC}
\mathbf{\Phi}^A \mathbf{\Phi}^B \mathbf{\Phi}^C \,,
\label{eq:24-superpotential}
\eeq
where $A,B,C$ denote $SU(5)_{\rm GUT}$ group indices and $d^{ABC}$ is the
fully symmetric $SU(5)$ structure constant. The other dimension-three operator $\mbox{Tr}[{\bf \Phi}{\bf \Phi}{\bf \Phi}] = d^{ABC}\mathbf{\Phi}^A \mathbf{\Phi}^B \mathbf{\Phi}^C/4$ is not independent. For simplicity, we ignore $SU(5)_{\rm GUT}$-breaking superpotential terms. Before we introduce SUSY-breaking effects, all 24 superfields have approximately degenerate masses of $M$. We neglect GUT-breaking running effects. The Yukawa $\lambda$ provides additional interactions for the superfields and can easily become strong because of the large multiplicity. Using the group property of $d^{ABC}d^{ABC^\prime} = (C_A^2 - 4)(C_A - 2 C_F)\delta^{CC^\prime}$ with $C_A = N_f = 5$ and $C_F = (N_f^2 -1)/2N_f$, the perturbativity bound for the Yukawa coupling is $\lambda < 16\pi/\sqrt{9\, (C_A^2 - 4)(C_A - 2 C_F)} \approx 8.2$.

For the SM-singlet component, ${\bf \Phi}_{(\bf{1},\bf{1})_{\bf 0}}$ and at loop level, it develops interactions with SM gauge bosons via triangle diagrams with other, SM-gauge-charged components of $\mathbf{\Phi}$ running in the loop.  In the limit
where the mass of the loop particles is large, the interactions with the three SM gauge fields are proportional to the changes of beta functions~\cite{Bertolini:2013via} and are calculated as
\beq
W = \sqrt{15}\frac{\lambda\, \alpha_s}{16\pi M} {\bf \Phi}^{(1,1)_0}\,
\mathbf{W}_G^{\alpha,a} \mathbf{W}_{G,\alpha}^a - \sqrt{15}
\frac{3\, \lambda\, \alpha_2}{32\,\pi\, M} {\bf \Phi}^{(1,1)_0}\,
\mathbf{W}_W^{\alpha,i} \mathbf{W}_{W,\alpha}^i - \sqrt{15}
\frac{\,\lambda\, \left(\frac{5}{3} \alpha_Y\right)}{32\,\pi\,M} {\bf \Phi}^{(1,1)_0}\,
\mathbf{W}_B^{\alpha} \mathbf{W}_{B,\alpha}  \,,
\eeq
with $a=1\,\cdots,8$ as the QCD gauge index and $i=1,2,3$ as the  $SU(2)_W$ gauge index. 
More precisely, a full loop calculation for the decay should be done, which introduces a kinematic correction factor.  This correction is small in the supersymmetric limit when all the $\mathbf{\Phi}$
states are degenerate.  

Given a superpotential of this form, we can expand in components to
determine the interactions of the scalar and pseudo-scalar components
of $\mathbf{\Phi}^{(1,1)_0}$ with gauge bosons.  We expand the superfield
as
\beq
\mathbf{\Phi}^{(1,1)_0} = \frac{1}{\sqrt{2}}\left[ \Phi^{(1,1)_0}_R + i \,\Phi^{(1,1)_0}_I \right] +
\sqrt{2}\theta\, \widetilde{\Phi}^{(1,1)_0} + \theta^2\,F_{\Phi^{(1,1)_0}} \,,
\eeq
where $\Phi^{(1,1)_0}_R$ is a scalar and $\Phi^{(1,1)_0}_I$ is a pseudo-scalar.
We then find, for example, that the couplings of the spin-zero particles
to gluons are given by
\beq
{\cal L} = \frac{\sqrt{15}\,\lambda\,\alpha_s}{16 \sqrt{2}\,\pi\,M}
\left[\Phi^{(1,1)_0}_R\,G^{\mu\nu\,a} G^a_{\mu\nu} \,+\, \Phi^{(1,1)_0}_I\, G^{\mu\nu\,a} \widetilde{G}^a_{\mu\nu}\right] \,.
\eeq
As discussed in the introduction, the couplings
of the scalar and pseudo-scalar are equal at the supersymmetric
level.  If SUSY breaking effects are small, they will form two nearly
degenerate resonances with nearly equal couplings to gauge bosons.  It
remains to verify that these breaking effects are indeed small.

Several SUSY breaking effects can alter the spectrum of states and their interactions.  In general, we can write soft potential as
\beq
{\cal L}_{\rm soft} =  (M_{\rm soft}^A)^2 |\Phi^A|^2 \, +\, \left[T_{\rm soft} \, \Phi^{(1,1)_0}\, + \, \frac{B_{\rm soft}^A}{2} \,(\Phi^A)^2 \,+ \, A_{\rm soft} \frac{d^{ABC}}{4} \Phi^A \Phi^B \Phi^C + {\rm h.c.}\right].
\eeq
We are most interested in effects that split the scalar and pseudo-scalar states in the chiral multiplets $\mathbf{\Phi}$.  The soft mass $M^A_{\rm soft}$ will not split these states.  The tadpole $T_{\rm soft}$ and $U(1)_\Phi$ breaking mass $B_{\rm soft}$ will both contribute to a mass splitting. The $A$-term $A_{\rm soft}$ can in principle split the couplings of the scalar and pseudo-scalar to gauge bosons, but requires an additional soft parameter insertion and will be subdominant in the limit that SUSY breaking effects are small.  

To determine the SUSY breaking parameters, we consider the minimal case of that the $\mathbf{\Phi}$
sector feels SUSY breaking only via SM gauge interactions. We are therefore working in the limit 
that the messengers of SUSY breaking do not couple to the $\mathbf{\Phi}$ sector as strongly as they couple to the MSSM sector. Assuming the gaugino masses are comparable, the largest SUSY breaking effects then affect the QCD-charged components of $\mathbf{\Phi}$, which we denote by $\mathbf{\Phi}^{(8,1)_0}$, $\mathbf{\Phi}^{(\overline{3},2)_{5/6}}$ and  $\mathbf{\Phi}^{(3,2)_{-5/6}}$.
The splitting of the bosons in these QCD-charged multiplets comes from a $B_{\rm soft}$ term.  At leading  order in $\log \Lambda$, we find
\beqa
M({\Phi}^{\rm QCD}_R) - M({\Phi}^{\rm QCD}_I) = \frac{2\, C_2(r)\, \alpha_s \, M_{\tilde{g}}}{\pi \, (M_{\tilde{g}}^2 - M^2)} \,\left(M_{\tilde{g}}^2 \, \log\frac{\Lambda^2}{M_{\tilde{g}}^2}  - M^2\, \log\frac{\Lambda^2}{M^2}\right)  \,.
\eeqa
Here, the quadratic Casimirs are $C_2(r) = 3$ for $\mathbf{\Phi}^{(8,1)_ 0}$ and $C_2(r) = 4/3$ for
$\mathbf{\Phi}^{(\overline{3},2)_{5/6}}$ and  $\mathbf{\Phi}^{(3,2)_{-5/6}}$. $\Lambda$ is the UV threshold that cuts off the divergent loop integral.
For $\frac{\alpha_s}{\pi}\, M_{\tilde{g}} \ll M$, we have approximately degenerate masses for the scalar and pseudo-scalar. 

The SUSY-breaking effects for the QCD-charged fields can feed into the SM-singlet field, $ {\bf \Phi}^{(1,1)_0}$, via the Yukawa coupling in Eq.~(\ref{eq:24-superpotential}).  At leading order in $\lambda$, both $T_{\rm soft}$ and $B_{\rm soft}$ contribute. They are generated by loops of QCD-charged scalars.  Both $M_{\rm soft}$ and $B_{\rm soft}$ terms for the QCD states can feed into this splitting, though the $M_{\rm soft}$ contributions are finite.  Note that the $M_{\rm soft}^2$ term for QCD charged states is of order
\beq
\left(M_{\rm soft}^r\right)^2 \sim \frac{C_2(r)\, \alpha_s}{4\pi} M_{\tilde{g}}^2 \log\frac{\Lambda^2}{M^2_{\tilde{g}}} \,.
\eeq
After summing all QCD-charged field contributions, the mass difference of the scalar and pseudo-scalar is calculated to be
\beqa
&& M\left[{\Phi}^{(1,1)_{0}}_R\right] - M\left[{\Phi}^{(1,1)_0}_I\right]  =  \frac{3 \lambda^2}{320\, \pi^2 M} \left[4 \left(3 B_{\rm soft}^{(3,2)_{-5/6}} - 8 B_{\rm soft}^{(8,1)_0}\right) \log \frac{\Lambda^2}{M^2} \right.\nonumber \\ && 
\left. \hspace{1cm} - 32 \left(M_{\rm soft}^{(8,1)_0}\right)^2  - 3 \left(M_{\rm soft}^{(3,2)_{-5/6}}\right)^2 - 15 B_{\rm soft}^{(3,2)_{-5/6}}\right]   
= {\cal O}\left( \frac{3\,\lambda^2\,\alpha_s}{10\pi^3}  \frac{M_{\tilde{g}}^2}{M}   \log{\frac{\Lambda^2}{M^2_{\tilde{g}}}  }\right) \,,
\label{eq:24MSSM-singlet-mass-split}
\eeqa
For $\alpha_s=0.09$, $\lambda=1$, $M=1$~TeV, $M_{\tilde{g}} = 2$~TeV and $\Lambda=5$~TeV, 
we have the mass splitting of the SM singlet scalars of ${\cal O}(10~\mbox{GeV})$, which is small compared to the overall meson mass.

\section{Mesons in SQCD with $N_f=N_c+1$}
\label{sec:model}
\subsection{General Considerations}
The second model we consider has superbump particles arising as composite particles from strong dynamics. Specifically, we consider supersymmetric QCD with an $SU(N_c)$ gauge group and $N_f$ vector-like quark flavors $\mathbf{Q}_i,~\widetilde{\mathbf{Q}}_i$. To introduce SM gauge interactions for this SQCD sector, we gauge all or a part of the diagonal flavor $SU(N_f)_V$ group. 
One natural embedding is to have $N_f = 5$ and identify the diagonal flavor symmetry as $SU(5)_{\rm GUT}$, which will be considered as a benchmark model. Another
possibility is to embed $SU(3)_C \times U(1)_Y$ into $SU(4)$ for $N_f
= 4$ and one could have more freedom to assign hypercharges for quarks. We first keep $N_f$ arbitrary and specify our results to the benchmark case with $N_f=5$ later.

\begin{table}[!tb]
\renewcommand{\arraystretch}{2.0}
\begin{center}
\begin{tabular}{c| c| c c c c}
\hline \hline
& $SU(N_c)$ & $SU(N_f)_L$ & $SU(N_f)_R$ & $U(1)_B$ & $U(1)_R$ \\
\hline \hline
$\mathbf{Q}$ & $N_c$ & $N_f$ & $1$ & $1$ & $\dfrac{1}{N_f}$ \\
$\widetilde{\mathbf{Q}}$ & $\overline{N}_c$ & $1$ & $\overline{N}_f$ & $-1$ & $\dfrac{1}{N_f}$ \\
\hline 
    ${\cal M}$ & 1 & $N_f$ & $\overline{N}_f$ & $0$ & $\dfrac{2}{N_f}$ \\
    ${\cal B}$ & 1 & $\overline{N}_f$ & $1$ & $N_c$ & $\dfrac{N_c}{N_f}$ \\
$\widetilde{\cal B}$ & 1&  $1$ & $N_f$ & $-N_c$ & $\dfrac{N_c}{N_f}$ \\
\hline \hline
\end{tabular}
\end{center}
  \caption{Superfield quark content in the UV (upper rows) and superfield meson and baryon content in the IR theory (lower rows) for $N_f = N_c + 1$ in SQCD.
 \label{tab:uv-fields}
 }
\end{table}
We show the gauge and global symmetries of the superfields in the UV in Table~\ref{tab:uv-fields}. Though axial $U(1)_A$ transformations of the quark superfields and $R$-symmetry-like transformations of the fermions are in general anomalous, there is a unique combination of these two transformations that is anomaly free. Choosing the normalization such that the $SU(N_c)$ gaugino has $U(1)_R$ charge one, we show the anomaly-free charges for the quark superfields in the last column of Table~\ref{tab:uv-fields}. Other than the gauge interactions for underlying quarks, we also add a quark mass matrix $m_0$ to the superpotential as
\beq
W_{\rm tree} \, = \, \mathbf{Q} \, m_0 \, \widetilde{\mathbf{Q}} \,.
\eeq

For $SU(N_c)$ SQCD models, the most phenomenologically interesting possibilities are
$N_f = N_c$ and $N_f = N_c + 1$, in which there is no additional unbroken gauge symmetries in the IR theory. For $N_f = N_c$, the low energy theory has  confinement with chiral symmetry breaking, while for $N_f = N_c + 1$ there is only confinement without chiral symmetry breaking~\cite{Intriligator:1995au}.  Both cases have mesons and baryons with an
unbroken vector-like gauged flavor group, though the parametric dependence in
these cases differs.  We focus on the case with $N_f = N_c + 1$, though similar results exist for $N_f = N_c$.  The two classes of models do not differ significantly in terms of their signatures.

For $N_f = N_c + 1$, the IR theory below a scale $\Lambda_{\rm conf}$ is confining, but does not
spontaneously break chiral symmetry, a scenario which is known as
s-confinement. The spectrum contains
$SU(N_c)$-singlet mesons, 
${\cal M}^{ij}\equiv \mathbf{Q}^i \mathbf{\widetilde{Q}}^j$, and
baryons, ${\cal B}_i\equiv \epsilon_{i\,i_1\cdots
  i_{N_c}}\mathbf{Q}^{i_1}\cdots \mathbf{Q}^{i_{N_c}}/N_c!$ and
$\widetilde{\cal B}_i\equiv \epsilon_{i\,i_1\cdots
  i_{N_c}}\mathbf{\widetilde{Q}}^{i_1}\cdots
\mathbf{\widetilde{Q}}^{i_{N_c}}/N_c!$. In terms of the global
symmetries, the particle content in the low energy theory transforms
as the lower part in Table~\ref{tab:uv-fields}.  The effective
superpotential for the low-energy theory is given by
\beqa
W_{\rm eff} = \frac{1}{\Lambda_{\rm conf}^{2N_f -3}}\left({\cal B}_i {\cal M}_{ij} \widetilde{\cal B}_j - \mbox{det}\,{\cal M} \right) \,+\, \mbox{Tr}(m_0\, {\cal M} ) \,.
\eeqa
The scale $\Lambda_{\rm conf}$ is typically taken to be the
scale at which the UV gauge coupling diverges, but for a practical
application of this model, we would like a more accurate estimate of
the coefficients of the various terms in the superpotential.  In
fact, one can apply a power counting scheme to rewrite the
superpotential in canonical normalization with $\mathcal{O}(1)$
coefficients.  Applying power counting in $4 \pi$ and $N_f$~\cite{Cohen:1997rt,Luty:1997fk,Cho:1998vc},
\beqa
\label{eq:ir-superpotential}
W_{\rm eff} = \frac{4\pi \lambda}{\sqrt{N_f}} \,\mathbf{B}_i
\mathbf{M}_{ij} \mathbf{\widetilde{B}}_j - \frac{(4\pi)^{N_c-1}}{\sqrt{N_c!} \,\Lambda^{N_c - 2} } \,\mbox{det}\,\mathbf{M} \,+\, \frac{\Lambda}{4\pi}\,\mbox{Tr}(m\, \mathbf{M} ) \,,
\eeqa
where $\lambda \sim 1$ and $\Lambda$ is roughly the scale at which
the effective theory breaks down.  The rescaled mass matrix $m$
corresponds to $m_0$ up to $\mathcal{O}(1)$ factors.  We normalize $\Lambda$ so as to
absorb the $\mathcal{O}(1)$ factor in front of the $\mbox{det}\,
\mathbf{M}$ term. 

Minimizing the
superpotential, one has the following supersymmetric
vacuum\footnote{We do not consider the meta-stable and
  supersymmetry-breaking vacuum, which has a higher value of vacuum
  energy~\cite{Intriligator:2006dd}.} 
\beqa
\langle \mathbf{B}_i \rangle = \langle \mathbf{\widetilde{B}}_i \rangle = 0 \,, 
\qquad 
\langle \mathbf{M} \rangle = \frac{\sqrt{N_c!}^{\frac{1}{N_c}} }{4\pi} \,
\Lambda^{\frac{N_c-1}{N_c}} \, ({\rm det}\,m)^{\frac{1}{N_c}} \, m^{-1}  \,.
\eeqa
From this vacuum, we find the following fermion mass matrices:
\beqa
m_{B,ij} = \frac{4\pi\lambda}{\sqrt{N_f}} \, \langle \mathbf{M} \rangle_{ij} \,,
\qquad m_{M,ijk\ell} = - \frac{(4\pi)^{N_c-1}}{\sqrt{N_c!}\,
  \Lambda^{N_c-2}} \,\left(\langle \mathbf{M}\rangle_{ji}^{-1} \, \langle
\mathbf{M}\rangle_{\ell k}^{-1}  - \langle \mathbf{M}\rangle_{jk}^{-1}
\, \langle
\mathbf{M}\rangle_{\ell i}^{-1}\right) \, {\rm det}\langle \mathbf{M} \rangle \,.
\eeqa
The bosons and fermions are degenerate in the absence of SUSY-breaking
effects.  

\subsection{Benchmark Example with $N_f = 5$}
To proceed further in studying the phenomenology of these models, we
now present some results regarding our specific benchmark model.  We
focus on $N_f = 5$ with a GUT embedding of the SM gauge group.
The representations of the quarks under the gauge group is shown in
Table~\ref{tab:fieldcontent-UV}.
\begin{table}[ht!]
\renewcommand{\arraystretch}{1.8}
\begin{center}
\begin{tabular}{c|c|ccc}
\hline\hline
         & $SU(N_c)$    &   $SU(3)_{\rm QCD}$  & $SU(2)_{W}$ & $U(1)_Y$   \\  \hline
         $\mathbf{Q}_3$        & $N_c$          & $\overline{3}$                &  1      & 1/3                \\ 
         $\mathbf{Q}_2$            & $N_c$          &  1               &  2      & -1/2    \\ \hline  
         $\mathbf{\widetilde{Q}}_3$  &  $\overline{N}_c$ &  3  &    1   & -1/3   \\ 
         $\mathbf{\widetilde{Q}}_2$  &  $\overline{N}_c$ &  1   &   2   &     1/2  \\ \hline  
 \hline
\end{tabular}
\end{center}
\caption{Field content of a supersymmetric QCD gauge theory with
  vector-like quark flavors, $\mathbf{Q}_{3, 2}$ and
  $\mathbf{\widetilde{Q}}_{3, 2}$, charged under SM gauge group for $N_f =
  5$.
\label{tab:fieldcontent-UV}}
\end{table}%
As seen above, the global symmetry of this SUSY QCD model is $SU(N_f)_L \times
SU(N_f)_R \times U(1)_B \times U(1)_R$. We then embed the SM gauge group as $SU(N_f)_L \times SU(N_f)_R \supset SU(N_f)_V \supset SU(3)_{\rm QCD} \times SU(2)_W \times U(1)_Y$. 
\begin{table}[ht!]
\renewcommand{\arraystretch}{1.8}
\begin{center}
\begin{tabular}{c|c}
\hline \hline
           &   $SU(3)_{\rm QCD}\times SU(2)_{W}\times U(1)_Y$  \\  \hline
${\cal M}$         &  $(8, 1)_0 + (\bar{3}, 2)_{5/6}
+ (3, 2)_{-5/6} + (1,3)_0 + (1, 1)_0+ (1, 1)_0$ \\ \hline
${\cal B}$        &  $(3, 1)_{-1/3} + (1, 2)_{1/2}$  \\ \hline 
${\tilde{\cal B}}$        &  $(\bar{3}, 1)_{1/3} + (1, 2)_{-1/2}$  \\ \hline  
 \hline
\end{tabular}
\end{center}
\caption{Field content under the SM gauge group in the low energy theory.
\label{tab:fieldcontent-IR}}
\end{table}%

Without loss of generality, we choose the mass matrix as $m =
\mbox{diag}\{m_3, m_3, m_3, m_2, m_2\}$.  The non-trivial meson VEV is
given by 
\beqa
\langle \mathbf{M} \rangle = \frac{24^{1/8} }{4\pi} \,
(m_2^2 \, m_3^3 \, \Lambda^3)^{1/4}  \, \mbox{diag}\{m_3^{-1}, m_3^{-1}, m_3^{-1}, m_2^{-1}, m_2^{-1} \}  \,.
\label{eq:SQCD-VEV}
\eeqa
We expand the meson into flavor components as
\beqa
\mathbf{M} \equiv \langle \mathbf{M} \rangle + \, \frac{\mathbf{H}^\prime}{\sqrt{5}}\,\mathbb{I}_5 + \sqrt{2}\,\mathbf{\Pi}^A T^A  \,.
\eeqa
where the traceless generators for $SU(5)$ are normalized such that
$\mbox{tr}[T^A T^B] = \frac{1}{2}\delta^{AB}$. The $SU(5)$ singlet,
$\mathbf{H}^\prime$, can be thought as the $\eta^\prime$ in the
SM. However, the $\mathbf{H}^\prime$ field could have a mass
comparable to other mesons in SQCD, which is different from the SM QCD sector. In fact, we will find that it generally
mixes with the SM singlet component, $\mathbf{\Pi}^{24}$,  of the 24 of $SU(5)$.  The
spectrum of the model can now be calculated.  In the absence of SUSY
breaking, the SM charged mesons all have simple expressions for their
masses.  For future convenience, we define the general meson mass scale as
\beq
M \equiv \frac{m_3^{3/4} \,\Lambda^{1/4}}{2^{3/8}\,3^{1/8}} \,,
\eeq
so that
\beqa
M_{{\bf \Pi}^{(1,3)_0}} = \left(\frac{m_2}{m_3}\right)^{3/2} \, M \,, \qquad 
M_{{\bf \Pi}^{(8,1)_0}} =  \left(\frac{m_2}{m_3}\right)^{-1/2} \, M \,, \qquad 
M_{{\bf \Pi}^{(3,2)_{-5/6}}} =  \left(\frac{m_2}{m_3}\right)^{1/2} \, M \,.
\eeqa
The baryons acquire a mass of 
\beqa
M_{{\bf B}^{(3, 1)_{-1/3} }} = \frac{2^{3/4} \, 3^{1/4}}{\sqrt{5}} \, \lambda \,  \left(\frac{m_2}{m_3}\right)^{1/2} \sqrt{\frac{\Lambda}{m_3}} \, M \,,  \qquad 
M_{{\bf B}^{(1, 2)_{1/2} }} = \frac{2^{3/4} \, 3^{1/4}}{\sqrt{5}} \, \lambda \,  \left(\frac{m_2}{m_3}\right)^{-1/2} \sqrt{\frac{\Lambda}{m_3}} \, M  \,,
\eeqa
which is larger than the pion masses in general.

The two singlets, $\mathbf{H}^\prime$ and $\mathbf{\Pi}^{24}$ with
$T^{24} = \mbox{diag}\{2,2,2,-3,-3\}/\sqrt{30}$, generally acquire a
mass mixing, though in the $m_2 \approx m_3$ approximation, the mixing
is suppressed.  The masses for $m_2 \ll m_3$ and for $m_2 = m_3$
are given by
\beqa
\renewcommand{\arraystretch}{1.5}
M_{{\bf \Pi}^{1^A} } \approx \left\{\begin{array}{l l} 2  \left(\frac{m_2}{m_3}\right)^{3/2} M (1 - \frac{3}{2} \frac{m_2^2}{m_3^2} )\,,
\\    M\,, 
    \end{array}\right. \quad
M_{{\bf \Pi}^{1^B} } \approx \left\{\begin{array}{l l} 2  \left(\frac{m_2}{m_3}\right)^{-1/2} M (1 + \frac{3}{2} \frac{m_2^2}{m_3^2} )\,, &
    m_2 \, \ll \, m_3 \,; \\
    4\,M\,, & m_2 \, = \, m_3 \,.\end{array}\right. 
    \label{eq:mass-formula-1A}
\eeqa
Note that in any spectra generated in subsequent sections, we use the
exact $m_2$ and $m_3$ dependence in our calculations.  After regulating the
singularity at $m_2 = 0$ in this expansion, the convergence remains
slow in $m_2 / m_3$ and the hierarchy between $m_2$ and $m_3$ is not sufficient to
neglect higher order terms for our purposes.  Nevertheless, the
formulas Eq.~\eqref{eq:mass-formula-1A} are valid at the $\sim 30\%$ level for
$m_2 / m_3 \sim 0.6$ and at the $\sim 10\%$ level for $m_2 / m_3 \sim 0.35$.
The SM singlet meson ${\bf \Pi}^{1^A}$ is 
lighter than ${\bf \Pi}^{1^B}$. The mixing matrix between the ${\bf
  H}^\prime, {\bf \Pi}^{24}$ basis and the mass basis ${\bf
  \Pi}^{1^A}, {\bf \Pi}^{1^B}$ is given by 
\beqa
\renewcommand{\arraystretch}{1.5}
\label{eq:mixing-matrix}
\begin{pmatrix} {\bf \Pi}^{1^A} \\ {\bf \Pi}^{1^B} \end{pmatrix} = \begin{pmatrix} \sin\theta
  &- \cos\theta \\ \cos\theta &
  \sin\theta \end{pmatrix} \begin{pmatrix} {\bf H}^\prime \\
  {\bf \Pi}^{24} \end{pmatrix},
  \qquad \quad  \mbox{with}\quad
  \sin\theta \approx \left\{\begin{array}{l l} \sqrt{\frac{2}{5}}\left(1 - \frac{3\,m_2}{2\,
      m_3}\right), &  m_2 \ll m_3\,; \\
    \frac{3 \sqrt{6}}{25} \frac{m_3 - m_2}{m_3}\,, & m_2 \approx
    m_3\,. \end{array}\right.
\eeqa
It is easy to see that when $m_2 = m_3$ the lighter state ${\bf \Pi}^{1^A}$ can be matched to ${\bf \Pi}^{24}$. In the other limit with $m_2 \ll m_3$, the lighter state ${\bf \Pi}^{1^A}$ can be identified as the meson associated with the $\mbox{diag}(0,0,0,1,1)$ generator.

\subsection{Meson Loop-generated Gauge Boson Interactions}
\label{sec:anomaly}
The neutral pion in the SM decays to two photons in a way that is
predicted by the mismatch of anomalies between the ultraviolet theory of
quarks and the low energy theory in which there are only bosonic meson
degrees of freedom.  The interaction can be treated as the pion
coupling to the anomalously non-vanishing divergence of the axial
current with a coefficient proportional to the mismatch in anomaly
coefficients.  Naively, such an interaction should arise in a SQCD
context as well, but matters are complicated by the fact that there
are fermions in the low energy: the superpartners of
mesons, as well as baryons.

't Hooft anomaly matching, together with holomorphy, provides a
non-trivial constraint on the potential low energy spectrum in the
SQCD as seen above.  The triangle anomalies for all unbroken
symmetries must be matched.  
In the absence of SUSY breaking and in the limit where the meson and
baryon masses are negligible, the anomalies for the non-Abelian $SU(N_f)_A$ axial
current and the non-anomalous $R$-symmetry current are
both matched between the UV and IR, such that, naively, there is no
loop-generated interaction between the mesons and a pair of gauge
bosons or gauginos.  In other words, couplings analogous the SM neutral pion
coupling to two photons vanish in this limit.\footnote{There is a mismatch of the anomaly of $U(1)_B\times SU(N_f)_A \times SU(N_f)_V$ in the UV and IR, which could have interesting phenomenological consequence if $U(1)_B$ is gauged.}

On the other hand, the fermionic mesons and baryons of the model
considered in Section \ref{sec:model} are not massless.  They in fact
generate a loop-level interaction between the meson and
gauge boson superfields after all, just like the simple example in Section~\ref{sec:24}. These one loop corrections to the gauge kinetic function are allowed as they are proportional to the beta function.  The structure of such couplings are analogous to the
loop couplings of the Higgs boson to gluons and photons, which are
known by the low energy theorem to be proportional to the change of beta
functions by integrating out heavy fields.

As in the {\small 24}MSSM, meson superfields $\mathbf{\Pi}$ will have meson and baryon loop induced interactions with gauge superfields.  For the general considerations below, we denote the loop meson or baryon by $\mathbf{\Phi}$ (and $\mathbf{\Phi}^c$ if it has complex gauge representation), where
$\mathbf{\Phi}$ may be either $\mathbf{M}$, $\mathbf{B}$ or
$\widetilde{\mathbf{B}}$.  We assume for simplicity that
$\mathbf{\Phi}$ is not self-conjugate, though all our results
generalize up to a factor of $1/2$.  We parametrize the interaction of $\mathbf{\Pi}$ with $\mathbf{\Phi}$ as $\lambda_\Phi \mathbf{\Pi} \mathbf{\Phi} \mathbf{\Phi}^c$.  In general, $\lambda_\Phi$ can be a tensor in gauge space if $\mathbf{\Pi}$ is part of a gauge multiplet, though we do not consider this case below.  We further work in the mass basis
for $\mathbf{\Pi}$ and $\mathbf{\Phi}$ for simplicity, in which the
states have mass $M_\Pi$ and $M_\Phi$ respectively.  In the limit
where $M_\Phi \gg M_\Pi$, which will be a good approximation for our purposes, the resulting interaction is given by 
\beq
W \supset - \frac{\lambda_\Phi\,b}{32 \pi^2 m_\Phi}
\mathbf{\Pi}\, \mathbf{W}^{A\,\alpha} \mathbf{W}^A_\alpha
 \,, \label{eq:loop-interaction}
\eeq
where $b$ is the beta
function contribution of $\mathbf{\Phi}$ and $\mathbf{\Phi}^c$.  
Note that the interaction in Eq.~\eqref{eq:loop-interaction} is presented
as merely the leading term assuming a flavor singlet pion.  A more general form of the interaction is
\beq
W \supset - \frac{1}{16 \pi^2}
\mathbf{W}^{A\,\alpha} \mathbf{W}^B_\alpha\, {\rm Tr}[ \log W_{\Phi\Phi^c} \, T^A T^B ]\,,
\eeq
where $W_{\Phi_i\Phi^c_j}$ is the derivative of the effective
superpotential with respect to $\mathbf{\Phi}_i$ and $\mathbf{\Phi}^c_j$ with $i,j$
flavor indices for $\mathbf{\Phi}$.

This interaction generates several
important interactions for the components of the superfields.  The
interactions of the scalar and pseudo-scalar with gauge bosons and
gauginos are all related by SUSY with relations protected by
holomorphy.  These relations force the decay rates of scalars
and pseudo-scalars to gauge bosons to be equal up to SUSY-breaking
effects.  On the other hand, the total widths of the scalars
and pseudo-scalars are not forced to be equal.  They generically differ due to
interactions of the scalar with the gauge $D$-term that are forbidden
for the pseudo-scalar.  Similarly, the scalar contraction of the
mesino with the gaugino can couple to a $D$-term, but the
pseudo-scalar contraction cannot.  Note that $D$-term-mediated decays for the scalars are only allowed if there are light gauge-charged matter fields.
\begin{table}[!tb]
\footnotesize
\renewcommand{\arraystretch}{1.2}
  \centering
  \begin{tabular}{c|c c c | c c c}
    \hline\hline
             & \multicolumn{3}{ c| }{$\mathbf{H}' \mathbf{W}^\alpha \mathbf{W}_\alpha$}    & \multicolumn{3}{ c}{$\mathbf{\Pi}^{24} \mathbf{W}^\alpha \mathbf{W}_\alpha$}   \\ \hline
    Field & $SU(3)_C$ & $SU(2)_W$ & $U(1)_Y$   & $SU(3)_C$ & $SU(2)_W$ & $U(1)_Y$  \\
    \hline
    $\mathbf{\Pi}_{(8,1)_0}$ & $3 (2 m_2 + m_3)$ & 0 & 0  
    & $\sqrt{6} (m_3 - 3 m_2)$ & 0 & 0    \\
    $\mathbf{\Pi}_{(1,3)_0}$ & 0 & $6 m_3$ & 0
    & 0 & $2 \sqrt{6} m_3$ & 0 \\
    $\mathbf{\Pi}_{(\bar{3},2)_{5/6}}$ & $2 (m_2 + 2 m_3)$ & $3 (m_2 +
    2 m_3)$ & $\frac{25}{3}(m_2 + 2 m_3)$ 
    & $\sqrt{\frac{2}{3} } (4 m_3 - 3 m_2)$ & $ \sqrt{\frac{3}{2}}(4 m_3 - 3 m_2)$ & $ \frac{25}{3\sqrt{6}}(4 m_3 - 3 m_2)$ \\
    $\mathbf{B}_{(\bar{3},1)_{1/3}}$ & $m_3$ & 0 & $\frac{2}{3} m_3$ 
    & $\sqrt{\frac{2}{3}} m_3$ & 0 &  $\frac{2}{3}\sqrt{\frac{2}{3}} m_3$\\
    $\mathbf{B}_{(1,2)_{-1/2}}$ & 0 & $m_2$ & $m_2$
    & 0 & $-\sqrt{\frac{3}{2}} m_2$ &  $-\sqrt{\frac{3}{2}}m_2$  \\
    \hline\hline
  \end{tabular}
  \caption{Couplings of $\mathbf{H}' \mathbf{W}^\alpha \mathbf{W}_\alpha$ and $\mathbf{\Pi}^{24} \mathbf{W}^\alpha \mathbf{W}_\alpha$
    generated by integrating out the gauge-charged 
    meson and baryon superfields.  We present the
    coupling as the coefficient of $1/[32 \sqrt{5} \pi^2\,\mbox{det}(\langle \mathbf{M} \rangle m)^{1/5}]$.}
  \label{tab:coefficients-hp}
\end{table}
We can proceed to write the components of
Eq.~\eqref{eq:loop-interaction} for the lighter SM-singlet meson
fields.  The contributions from the various gauge-charged superfields
to the coefficients $\lambda_\Phi b / m_\Phi$ for 
the gauge basis mesons are shown in Tables~\ref{tab:coefficients-hp}.   The masses of the mesons and baryons have already been
calculated in Section \ref{sec:model}.  We use
Eq.~\eqref{eq:mixing-matrix} to rotate the meson fields to the mass
basis, focus on the $\Pi^{1^A}$ meson and expand it as $\Pi^{1^A} = (\Pi^{1^A}_R +
i \Pi^{1^A}_I) / \sqrt{2}$.  A full expansion of superfields
into components is presented in Appendix~\ref{appendix:other-couplings}.  Here, we focus on the
phenomenologically interesting decays of scalars and pseudo-scalars to
gauge bosons. For example, the resulting $\Pi^{24}$-gauge boson
interactions are
\beqa
\label{eq:anomaly-meson}
{\cal L} & \supset & \frac{1}{32 \sqrt{15} \pi^2
  \mbox{det}(\langle \mathbf{M} \rangle m)^{1/5}} \, \left\{ 
  g_s^2\, (8\,m_3 - 12\,m_2) \,
(\Pi^{24, R} \,G^{a\,\mu\nu} G^a_{\mu\nu} + \Pi^{24, I}\,  G^{a\,\mu\nu}
\widetilde{G}^a_{\mu\nu})  \right. \nonumber \\
&&\left. \hspace{3.9cm} \,+\,   g^2\, (12\,m_3 - 6\,m_2) \,
(\Pi^{24, R}\, W^{i\,\mu\nu} W^i_{\mu\nu}  + \Pi^{24, I} \, W^{i\,\mu\nu}
\widetilde{W}^i_{\mu\nu})  \right. \nonumber \\
&& \left. \hspace{3.5cm} \,+\,   g^2_Y\, (\frac{52}{3}\,m_3 - 14\,m_2) \,
(\Pi^{24, R} \, B^{\mu\nu} B_{\mu\nu}  + \Pi^{24, I}  \, B^{\mu\nu} \widetilde{B}_{\mu\nu}
) 
\right\} \,,
\eeqa
where $\widetilde{F}^{\mu\nu} =
\frac{1}{2}\,\epsilon^{\mu\nu\rho\sigma} F_{\rho\sigma}$.   
Note that the couplings to
the scalar and pseudo-scalar are equal at this level. For equal
couplings as above, as well as equal masses, the partial widths of the scalar
and pseudo-scalar to gauge bosons are equal.  For instance, the partial widths to
gluons and photons are given by
\beqa
\Gamma(\Pi^{1^A}_{R,I} \to gg) & = & \frac{\alpha_s^2\,M_{\Pi_1^A}^3}{20 \cdot
  2^{3/4} 3^{1/4} \pi m_2 (\Lambda m_3)^{3/2}}\, \left[8 (m_2 + m_3)
\sin\theta - 4 \sqrt{\frac{2}{3}} (2\,m_3 - 3\,m_2) \cos\theta \right]^2 \,,  \label{eq:1A-decay-formula}\\ 
\Gamma(\Pi^{1^A}_{R,I} \to \gamma\gamma) & = & \frac{\alpha^2\,M_{\Pi_1^A}^3}{160 \cdot
  2^{3/4} 3^{1/4} \pi m_2 (\Lambda m_3)^{3/2}}\, \left[\frac{8}{3} (5\,
  m_2 + 11\,m_3) \sin\theta - \frac{4}{3} \sqrt{\frac{2}{3}} (22\,m_3 -
  15\,m_2) \cos\theta\right]^2\,, \nonumber
\eeqa
respectively.  Additional decays are discussed in Appendix~\ref{appendix:other-couplings}.  There is an additional decay mode for the real scalar via its
$D$-term coupling, though this decay is a small contribution to its
total width and phenomenologically challenging to observe.

The decays of both the $\Pi^{1^A}$ scalar and pseudo-scalar follow a familiar
pattern to the non-SUSY case, with dominant decays to pairs of
gluons.  Though SUSY-breaking effects can split the total widths of
the two spin-zero states, their branching fractions remain roughly the
same, up to the decay via $D$-term couplings mentioned above, which is
subdominant.  The branching fractions in two different limits are
presented in Table~\ref{tab:branching} for the particle mass
sufficiently above the electroweak gauge boson masses.
\begin{table}[htb!]
\renewcommand{\arraystretch}{1.7}
  \centering
  \begin{tabular}{c|c c c c c }
    \hline \hline
    Mode & $gg$ & $\gamma\gamma$ & $ZZ$ & $Z\gamma$ & $WW$  \\
    \hline
    Branching ratio (for $m_2 \sim m_3$) & $0.90$ & $4.6 \times 10^{-3}$ & $0.021$ & $8.3 \times 10^{-3}$
    & $0.066$  \\
    Branching ratio (for $m_2 = m_3/3$) & $0.17$ & $0.058$ & $0.18$ & $0.051$
    & $0.53$  \\
    \hline \hline
  \end{tabular}
  \caption{Decaying branching ratios for the scalar and pseudo-scalar particles in $\Pi^{1^A}$.  For the scalar field, there is an additional decay to two Higgs bosons, which has a branching ratio less than $10^{-4}$ for a 1 TeV particle mass. Here, we also ignore the potential decaying channels into heavy Higgs fields and lightest supersymmetric particles, by assuming kinematically inaccessibility. }  \label{tab:branching}
\end{table}
The phenomenology of the other mesons and mesinos is discussed further
in subsequent sections.

For the adjoint meson superfields, $\mathbf{\Pi}^{(8, 1)_0}$ and $\mathbf{\Pi}^{(1, 3)_0}$, a similar analysis for the SM-singlet superfields applies and has the interactions in the gauge kinetic function as $d^{abc}\,\mathbf{\Pi}^{(8,1)_0,a}\, \mathbf{W}_G^{b\,\alpha} \mathbf{W}^c_{G,\alpha}$, $\mathbf{\Pi}^{(8,1)_0,a}\, \mathbf{W}_G^{a\,\alpha} \mathbf{W}_{B,\alpha}$, and $\mathbf{\Pi}^i_{(1,3)_0}\, \mathbf{W}_W^{i\,\alpha} \mathbf{W}_{B,\alpha}$, which mediate interesting decay channels for mesons and mesinos.

\subsection{Supersymmetry Breaking Effects}
\label{sec:susy-breaking-general}
There are two sources of SUSY breaking effects: soft masses for the underlying quark superfields if the SQCD sector directly interacts with the hidden SUSY-breaking sector and the SM gauge interaction mediated SUSY breaking from the soft terms in the MSSM. We study both effects below. 
\subsubsection{Directly Mediated Supersymmetry Breaking Effects}
\label{sec:susy-breaking}
For the first source, in order to study SUSY-breaking effects, we parametrize potential effects by a SUSY-breaking spurion $\mathbf{X} = \theta^2 F$, where $F$ is a constant.  This spurion can either be inserted
linearly in the superpotential or quadratically in the K\"ahler
potential.  We are primarily interested in two SUSY breaking effects.
The first will split the scalar and pseudo-scalar components of the
meson and baryon superfields.  If the splitting is small, as it will
turn out to be for the parameter space of interest, then the two
scalars will have nearly degenerate mass.  Phenomenologically, this
results in two nearby resonances, which we call ``{\it superbumps}".
The second breaking effect is a 
splitting of the couplings of the singlet mesons to SM gauge bosons.
This effect, along with the aforementioned mass splitting, will both
contribute to the bumps having slightly different heights.  The
splitting in the couplings arises entirely due to SUSY-breaking
effects that cause the mixing angles for the scalar and pseudo-scalar
mesons to differ.  Throughout, we must also assume that SUSY breaking
effects are small compared to some combination of the confinement
scale and the SUSY-conserving quark superfield masses.  We will make
the structure of this approximation clearer below.

The mass splitting between the scalar and pseudo-scalar is generated
by a superpotential contribution as 
\beqa
W \, \supset \, \kappa_3 \, \mathbf{X} \, \mathbf{Q}_3 \,
\mathbf{\widetilde{Q}}_3 \, + \, \kappa_2 \, \mathbf{X} \, \mathbf{Q}_2 \, \mathbf{\widetilde{Q}}_2\,.
\eeqa
We define $\kappa =
\mbox{diag}(\kappa_3,\kappa_3,\kappa_3,\kappa_2,\kappa_2)$ and take $\kappa_2$ and $\kappa_3$ as real numbers.\footnote{In general, both $\kappa_2$ and $\kappa_3$ can be complex numbers and lead to CP-violating effects in our SQCD sector (see Ref.~\cite{Draper:2016fsr} for related discussion about the subsequent SM strong CP problem). }  For
simplicity, we work with $\kappa_3 = \kappa_2$.  We expect the
coefficients $\kappa_i$ to be order of
\beq
\kappa_i \sim \frac{m_{i}}{M_{\rm messenger}}\,,
\eeq
where $M_{\rm messenger}$ is the scale of the messengers that couple the
new quarks to the SUSY-breaking sector.
In the low energy theory, we expect to generate a term 
\beqa
\label{eq:breaking-super}
W \, \supset \, \frac{\Lambda}{4\pi} \mathbf{X} \, \mbox{Tr}(\kappa \,\mathbf{M})\,.
\eeqa
Such a term in the superpotential leads to a tadpole term for the
meson scalar component.  After shifting the meson vacuum and
diagonalizing the resulting mass matrix, the physical spectrum will
have a split between the masses of scalar and pseudo-scalar mesons.
In calculating the spectrum, we make the useful approximation that the
shift in the vacuum is small,
\beqa
F \ll m_3^{7/4}\, \Lambda^{1/4} \,.
\eeqa
In the limit of extremely small $m_2 \ll m_3$, this approximation can break
down before $F \sim m_3^{7/4} \Lambda^{1/4}$, but we never consider such
a toxic situation in our study.  Given the approximations outlined
above, we have SM-charged meson mass splittings as given in Table
\ref{tab:splitting}.  
\begin{table}[!tb]
\renewcommand{\arraystretch}{1.9}
\centering
\begin{tabular}{c| c |c| c |c |c}
\hline \hline
Meson &   $\Pi_{(1,3)_0}$ &  $\Pi_{(8,1)_0}$  & $\Pi_{(3,2)_{-5/6}}$  & $\Pi^{1^A}$ & $\Pi^{1^B}$ \\ \hline 
$\frac{M_{\rm scalar} - M_{\rm pseudo}}{\Delta M_{\rm split} }$ &
$\dfrac{2m_3 - m_2}{m_2}$ & $\dfrac{5m_2 - 2 m_3}{3m_2}$ & $\dfrac{2m_3 + m_2}{3m_2}$ & 
$\begin{array}{l l} 1, & m_2 =
    m_3 \\ \dfrac{2\,m_3}{m_2}, & m_2 \ll m_3\end{array}$  &
$\begin{array}{l l} -1, & m_2=
    m_3 \\ \dfrac{2\,m_3}{3\,m_2}, & m_2 \ll m_3\end{array}$ \\
\hline \hline
\end{tabular}
\caption{SUSY-breaking induced mass splittings between scalar and
  pseudo-scalar mesons.  We define $\Delta M_{\rm split} =3\, \kappa_3\, F / 4 m_3$.}  \label{tab:splitting}
\end{table}

For the mixing of states of $H^\prime$ and $\Pi^{24}$, the scalars and pseudo-scalars have different corrections to the rotation angles from Eq.~(\ref{eq:mixing-matrix}) due to SUSY-breaking effects. We find that the rotation angle difference is
\beqa
\renewcommand{\arraystretch}{1.9}
\sin\theta_R - \sin\theta_I \approx 
 \left\{
 \begin{array}{l l} \frac{3^{9/8} \,\kappa_3\,F \sqrt{m_2}} {2^{1/8} \,5^{1/2} \, m_3^{9/4} \, \Lambda^{1/4}} , &  \qquad m_2 \ll m_3\,; \\
    \frac{2^{15/8} \, 3^{5/8}\, F\,(m_3 - m_2) }{25\,m_3^{11/4}\,\Lambda^{1/4} } \,, & \qquad m_2 \approx
    m_3\,. \end{array}\right.
\eeqa
which slightly changes the production cross sections for the
scalar and pseudo-scalar at colliders. 

We further note that there are additional SUSY breaking effects due to
K\"ahler soft masses for the UV quarks, which leads to additional
Hermitian mass terms for the mesons in the IR.  Such mass terms cannot
split the scalar and pseudo-scalar components of the meson multiplet
and are thus of lesser phenomenological interest.  For simplicity, we
have neglected them, though they can easily be incorporated provided that 
they are sufficiently small as to be perturbations on our SQCD
analysis.  One potentially interesting consequence of such
contributions is that, if the meson soft mass contributions are
negative, they can lead to a situation where the scalar and
pseudo-scalar mesons are lighter than the fermionic mesinos.  This
kind of scenario is forbidden by experimental searches in the MSSM,
but should be possible generically.  The spin-zero components could then be
the first ones to show up in LHC and other collider searches.

\subsubsection{Supersymmetry-breaking Effects from the MSSM}
\label{sec:susy-breaking-MSSM}
Even the SQCD sector has small or vanishing coupling to the sector that mediates MSSM SUSY breaking, the SM gauge interactions themselves  mediate  SUSY-breaking effects into the SQCD sector as in the simple {\small 24}MSSM in Section~\ref{sec:24}. The MSSM gaugino soft mass can generate soft masses for the quark fields in SQCD at loop level. Neglecting the small $SU(2)_W\times U(1)_Y$ gauge interactions, the spurion-based superpotential in the UV has the form
\beqa
W \supset  \mathbf{Q}\, \mathbf{X}_g\, \mathbf{\widetilde{Q}} \,.
\eeqa
Here, the spurion field is written as a $5\times 5$ matrix with $\mathbf{X}_g = \mbox{diag}\{F_g, F_g, F_g, 0 ,0\}$ with 
\beqa
F_g \sim \frac{\alpha_s}{\pi} \, m_{3, 0} \,  M_{\tilde{g}}\,  \log{(\Lambda^2_{\rm UV}/M_{\tilde{g}}^2)} \,,
\eeqa
by ignoring additional terms of ${\cal O}(m^2_{3,0}/M^2_{\tilde{g}})$. The scale $\Lambda_{\rm UV}$ is related to the SUSY-breaking scale. In the low energy theory, we expect terms of 
\beqa
W \supset \eta_1 \frac{\Lambda}{4\pi} \mbox{Tr}( \mathbf{X}_g \,\mathbf{M} ) \,-\, 
\eta_2\,\frac{(4\pi)^3}{\Lambda^3}\,\mathbf{X}_{\rm instanton}\, f^{aBD}f^{aCE} \, \frac{\partial \,\mbox{det}\,\mathbf{M}}{\partial \mathbf{M}_B\,  \partial \mathbf{M}_C}\, \mathbf{M}_D \, \mathbf{M}_E \,+\, \cdots\,,
\label{eq:MSSM-softmass} 
\eeqa
where $a=1,2,\cdots,8$ are QCD gluon indices and $B,C,D,E$ are $SU(5)$ adjoint indices which enumerate the QCD-charged mesons $(8,1)_0$, $(\overline{3},2)_{5/6}$ and $(3,2)_{-5/6}$; $\eta_{1,2} > 0$ are order-one numbers.  $\mathbf{X}_g$ is carried over from the spurion in the quark theory by substituting $m_{3,0}$ by $m_3$. A new spurion, $\mathbf{X}_{\rm instanton}$, is generated by gluino-mediated one-loop diagram together with an instanton generated vertex. Its $F$-term, $\mathbf{X}_{\rm instanton} =\theta^2 F_{\rm instanton}$, is estimated to be 
\beqa
F_{\rm instanton} \sim \frac{\alpha_s}{\pi}\, \Lambda \, M_{\tilde{g}}\, \log{(\Lambda^2_{\rm UV}/M_{\tilde{g}}^2)} \,.
\eeqa

As in the calculation in Section~\ref{sec:susy-breaking}, the first tadpole-like term shifts the VEV's of mesons and contributes to both SM-singlets and SM-charged meson mass splittings. The second term in Eq.~(\ref{eq:MSSM-softmass}) only generates mass splittings for QCD-charged mesons. For the SM-singlet mesons, we have mass splittings as
\beqa
\renewcommand{\arraystretch}{1.9}
\Delta M\left({\Pi_{1_A}}\right) \equiv M_{\Pi_R^{1_A}} -
M_{\Pi_I^{1_A}} = 
\left\{\begin{array}{l l}   
\eta_1\,\dfrac{\alpha_s}{20\,\pi}\,M_{\tilde{g}}\, \log{(\Lambda^2_{\rm UV}/M_{\tilde{g}}^2)} \,, & \quad m_2 = m_3 \,; \\
-\eta_1\,\dfrac{3\,\alpha_s}{2\,\pi}\,\left( \dfrac{m_2}{m_3} \right)^{3/2}\,M_{\tilde{g}}\, \log{(\Lambda^2_{\rm UV}/M_{\tilde{g}}^2)}\,,  & \quad m_2 \ll m_3 \,.\end{array}  \right.
\label{eq:singlet-split-MSSM}
\eeqa
We find that the mass splitting is 5 GeV for $m_2 = m_3$,  $\eta_1=1$, $\alpha_s=0.09$, $M_{\tilde{g}}=2$~TeV and $\Lambda_{\rm UV} = 5$~TeV.  

For the color-octet mesons and after substituting the meson VEV's from Eq.~(\ref{eq:SQCD-VEV}), we have the mass splitting of QCD-charged mesons as 
\beqa
&&\Delta M[\Pi^{(8,1)_0}] = M_{\Pi_R^{(8,1)_0}} - M_{\Pi_I^{(8,1)_0}}  =
\eta_1\,\frac{5\, F_g}{2\, m_3} +
 \eta_2\, \frac{\sqrt{3}\,F_{\rm instanton}}{\sqrt{2}\,\Lambda} \nonumber \\
 & &  \hspace{2cm} =
\left( \eta_1\, \frac{5\,}{2}  \,+\,
  \eta_2\,\sqrt{\frac{3}{2}}\right)\, \frac{\alpha_s}{\pi}\,M_{\tilde{g}}\, \log{(\Lambda^2_{\rm UV}/M_{\tilde{g}}^2)}  \,.
  \label{color-split-MSSM}
\eeqa
For $\eta_1 = \eta_2 = 1$, the mass splitting is 390 GeV for $m_2 = m_3$,  $\alpha_s=0.09$, $M_{\tilde{g}}=2$~TeV and $\Lambda_{\rm UV} = 5$~TeV. A similar formula applies to $(\overline{3},2)_{5/6}$ and $(3,2)_{-5/6}$ with a different order-one number.  

\subsection{Additional Meson and Mesino Interactions}
\label{sec:meson-pheno}
We begin our study of meson and mesino phenomenology by
discussing meson and mesino decays.  At the level discussed so far,
the singlet states can decay via the anomaly induced interaction. The color octet and electroweak triplet states can also decay via anomaly induced interactions. The complex SM charged states are exactly stable at the level discussed so far, which is phenomenologically untenable.  Fortunately, there are
several non-renormalizable operators that we either expect to be
induced or that can be trivially added to the model while respecting
the continuous symmetries.  Before discussing their decays, we first point out additional potential decay modes for the adjoint charged particles 
$\mathbf{\Pi}_{(8,1)_0}$ and $\mathbf{\Pi}_{(1,3)_0}$. We expect transition anapole-charge radius operators to be
generated in the K\"ahler potential:
\beq
K \supset  -i \,c^s_{a-c}\,\frac{g_s}{\Lambda^2} \,\mathbf{H}^{\prime\dagger} D^\alpha
\mathbf{\Pi}^{(8,1)_0,a} \mathbf{W}^a_{G,\alpha} -i\,c^w_{a-c}\,\frac{g_2}{\Lambda^2} \,\mathbf{H}^{\prime\dagger} D^\alpha
\mathbf{\Pi}^{(1,3)_0,i} \mathbf{W}^i_{W,\alpha} + {\rm h.c.}\,,
\label{eq:anapole-charge-radius}
\eeq
where $D^\alpha$ is the SUSY covariant derivative and $c^{s,w}_{a-c}$
are numbers of order unity. This operator generates a
transition charge radius for scalars and a transition anapole
moment for fermions, allowing them to decay via the singlet
scalars, which may be either on-shell or off-shell.   In addition, if
there is a light SM gaugino and the fermion-scalar meson mass splitting is
sufficiently large, this term introduces decays to the light gaugino.
The operators in Eq.~\eqref{eq:anapole-charge-radius} can be expanded in
components.  For example, the first term generates scalar-gauge boson
interactions as 
\beqa
{\cal L} & \supset & \frac{1}{2} \,c^s_{a-c}\, \sin\theta\, \frac{g_s}{\Lambda^2}
\, \left\{ \left[\partial_\mu \Pi^{1^A}_R \, \partial_\nu \Pi^{(8,1)_0,a}_R \,
+ \, \partial_\mu \Pi^{1^A}_I \, \partial_\nu \Pi^{(8,1)_0,a}_I \right] \,
G^{a\mu\nu} \right. \nonumber  \\
~ & ~ & \left.\hspace{2.5cm} +\, \left[\partial_\mu \Pi^{1^A}_I \, \partial_\nu \Pi^{(8,1)_0,a}_{R}\,
- \,\partial_\mu \Pi^{1^A}_R \,\partial_\nu \Pi^{(8,1)_0,a}_{I}\right] \,\widetilde{G}^{a\mu\nu}\right\},
\eeqa
assuming that the coefficient $c^s_{a-c}$ is real, which corresponds
to the CP-conserving limit.
Note that the coupling to $G$ and $\widetilde{G}$ vanishes for identical scalars:
it only exists for a transition charge radius operator.

The situation for the $\mathbf{\Pi}_{(3,2)_{-5/6}}$ particle is less simple.  If we
do not write  an operator to decay this particle, it will not be
generated as there are residual accidental discrete symmetries in
the model as it has been written so far under which the new quarks are
odd.  For example, there is a $Z_2$ under which $\mathbf{Q}_2$ and
$\mathbf{\widetilde{Q}}_2$ are odd, but $\mathbf{Q}_3$ and
$\mathbf{\widetilde{Q}}_3$ are even.  There are simple K\"ahler terms
that can be written which will decay these particles and at the same time break the
discrete symmetries.  For example, we can introduce
\beq
K \supset \frac{\Lambda}{4\pi  M_{\rm UV}^2} \mathbf{\Pi}^{(\overline{3},2)_{5/6}}
\mathbf{u}^c \mathbf{q}^\dagger +
\mbox{h.c}  \,,
\eeq
where $\mathbf{u}^c$ and $\mathbf{q}$ are MSSM quark superfields of
arbitrary flavors with gauge index contracted and
$M_{\rm UV}$ is the scale of the corresponding dimension-six UV
operator.  There exist similar higher-dimension operators for the
adjoint meson superfields coupling to two MSSM quark superfields like
$\Lambda\, \mathbf{\Pi}^{(8,1)_0,a}(\mathbf{q}^\dagger t^a
\mathbf{q})/ 4\pi M_{\rm UV}^2$. For a sufficiently high UV scale
$M_{\rm UV} \gg \Lambda$, this interaction is less important than the
one in Eq.~(\ref{eq:anapole-charge-radius}) and will be ignored for
the adjoint meson phenomenological properties.

\section{Phenomenology of Superbumps}
\label{sec:pheno}

\subsection{Superbumps for a High Confinement Scale}
\label{sec:high-scale}
Even when the confinement scale or the low energy cut off scale,
$\Lambda$, is very high, we could still have light meson and baryon
superfields because their mass is suppressed by the bare masses $m_2$
and $m_3$. A priori, there is no simple argument to determine the
actual scale for $\Lambda$, except that it should be above the meson
field masses to justify our SQCD analysis. On the other hand, the
phenomenological consequence for different values of $\Lambda$ would
be different. For instance, using the formula in
Eq.~(\ref{eq:1A-decay-formula}) we have the decay lengths of the
singlet meson field as
\beqa
c\,\tau_0(\Pi^{1^A}) \approx 20~\mu{\mbox m}\,\times\,
\left(\frac{\Lambda}{10^8~\mbox{TeV}}\right)^{4/3}
\left(\frac{1~\mbox{TeV}}{M_{\Pi^{1^A}}}\right)^{7/3} \,,
\eeqa
in the limit of $m_2=m_3$. In other words, for a sufficiently high confinement scale, there could
be a displaced vertex signature.  Such a scenario would significantly
reduce the background for a superbump search. Pair production of SM charged states will have striking signatures if their masses are within the collider reach. 

\begin{figure}[!tb]
  \centering{
  \includegraphics[width=0.45\textwidth]{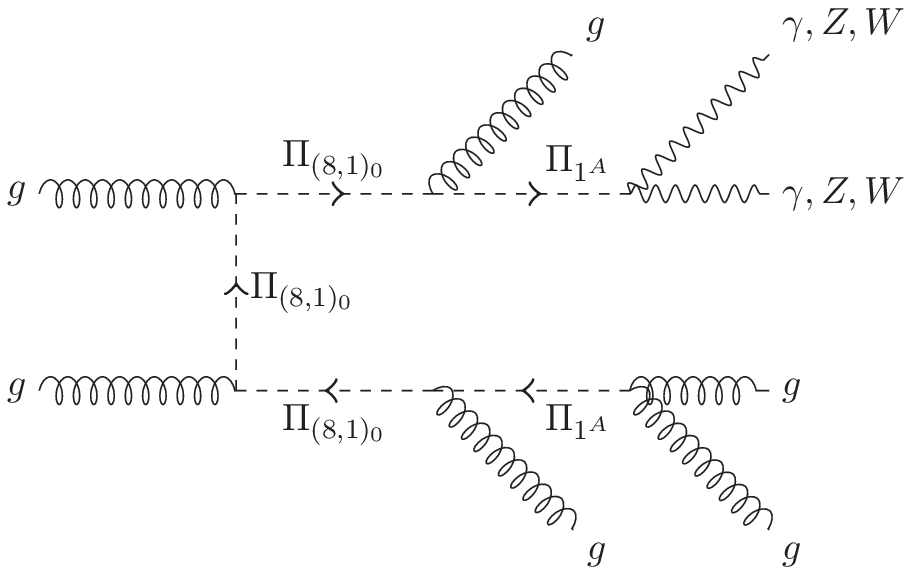} \hspace{0.5cm}
  \includegraphics[width=0.5\textwidth]{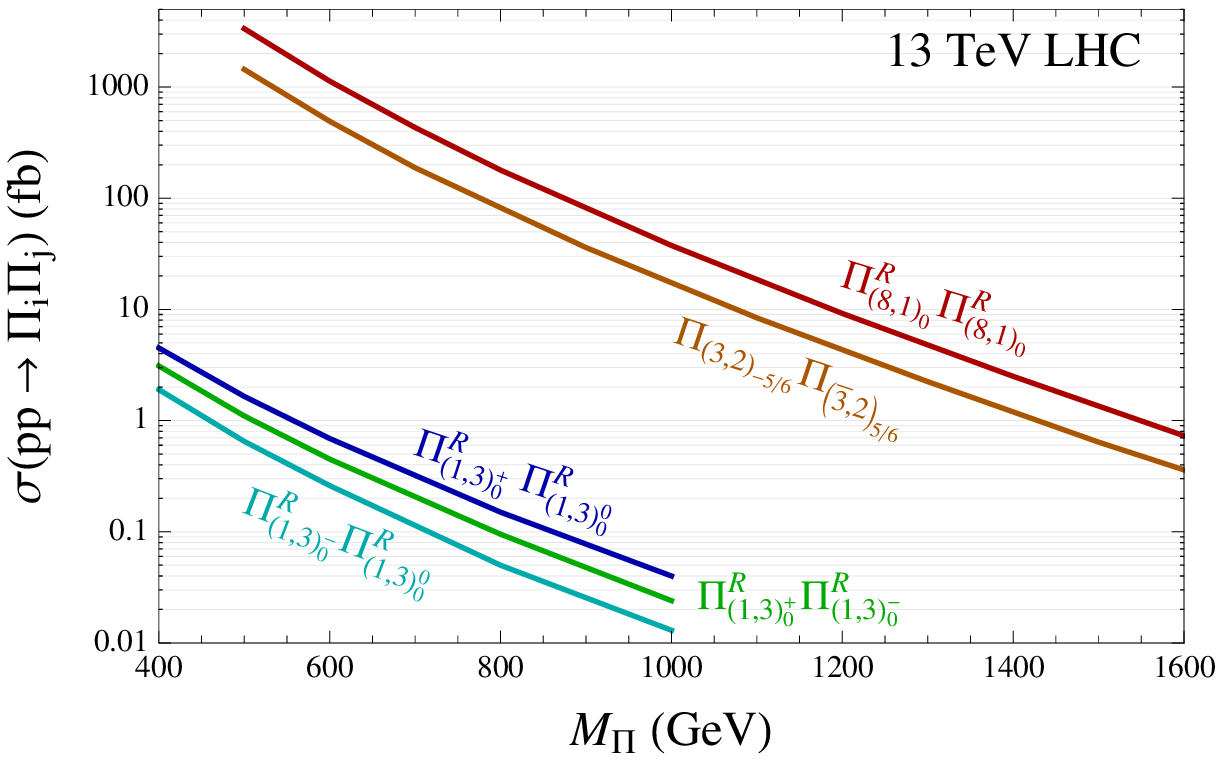}}
  \caption{Left panel: potential Feynman diagram for the production and decays of high confinement scale superbumps at a hadron collider. The scalar and pseudo-scalar of $\Pi^{(8,1)_0}$ have similar production cross sections and similar decay chains. Right panel: tree-level pair production cross sections for real scalar meson fields for adjoint representation mesons and complex scalar meson fields for the $(3,2)_{-5/6}$ and $(\overline{3},2)_{5/6}$ mesons,  from SM gauge interactions. The pseudo-scalar mesons have the same production cross sections as the scalar mesons. Here,  $\Pi_R^{(1,3)^{\pm,0}_0}$ represent the electric charged and neutral components of the weak triplet meson.  $\Pi^{(3,2)_{- 5/6}}$ and $\Pi^{(\overline{3},2)_{5/6}}$ denote complex scalars such that the line represents the combined cross-section for all states in these representations.}
  \label{fig:pair-production}
\end{figure}

For a moderately high confinement scale with $\Lambda \gtrsim 4\pi\times10$~TeV, the dominant production mechanism is likely to be pair production of SM-charged mesons, since  resonant production of a single meson field is suppressed by $1/\Lambda^{4/3}$ for fixed singlet masses. For instance and in the left panel of Fig.\ 
\ref{fig:pair-production}, we show the pair production Feynman diagram for the color-octet mesons at a hadron collider. In the right panel, we show the pair-production cross sections of various mesons at the 13 TeV LHC. After they are produced, the octets could then decay via the anapole-charge radius operator in Eq.~\eqref{eq:anapole-charge-radius} to the SM singlet meson, which then decays to two SM gauge bosons via the anomaly-mediated coupling. As a result, one could search for a pair of $3j + 3j$, $j2\gamma + 3j$, $j2Z + 3j$, $\cdots$, resonances.  The color octet scalars can also decay via an anomaly-mediated operator to two gluons with a coupling proportional to $d^{abc}$ of $SU(3)_c$~\cite{Bai:2010dj}. The corresponding signature is a pair of dijet resonances. The current mass constraint on pair produced scalar color octet decaying to jets is 700 GeV~\cite{Khachatryan:2014lpa}.  The color triplet scalars appear as diquarks and will also appear as pair produced dijet resonances, with a less stringent mass constraint of 600 GeV due to the slightly smaller color factor.

The tree-level pair production cross section for a scalar or pseudo-scalar color-octet meson with a mass of 1 TeV at the 13 TeV LHC is
around 35 fb. After multiplying it by the singlet digluon and diphoton
branching ratios, we have the interesting signature of $pp \rightarrow
2\,\Pi^{(8,1)_0}_{R, I} \rightarrow (j2\gamma)(3j)$ with a cross
section of 0.3 fb, using the branching ratios for $m_2\approx m_3$. The SM background is unlikely to mimic this
signature because of two equal-mass resonances on each decay chain. The LHC
Run II should therefore have a good chance to discover the color-octet
states in our model.

After the discovery of color-octet mesons, it is interesting to study CP properties. The simplest approach is to
consider the singlet meson decays of $\Pi_{R,I}^{1^A} \rightarrow (Z
\rightarrow \ell^+ \ell^-) (Z \rightarrow \ell^+ \ell^-)$ and study
the signal distribution of the angle between the two $Z$ boson decay
planes, as in the analysis for determining the SM Higgs CP
property~\cite{Gao:2010qx,Chatrchyan:2012jja,Aad:2015mxa}. Taking into
account of the $Z$ boson leptonic decay branching ratio, the final
signature production cross section of $pp \rightarrow
2\,\Pi_{(8,1)_0}^{R, I} \rightarrow (j 2\ell^+ 2\ell^-)(3j)$ is 0.003
fb at the 13 TeV LHC, which means a high-luminosity LHC run is
required to uncover the CP properties. One could also consider the
case with one leptonic $Z$ and one hadronic $Z$, with the final state
as $(j 2\ell^+ 2j)(3j)$ and 0.07 fb at the 13 TeV LHC. This final
state is less clean than the final state with four leptons, so a more
careful collider study is needed to test its feasibility.

For the lightest electroweak triplet mesons with a mass of a few hundred
GeV, the electric-charged component can decay to $W^\pm \gamma(Z)$, while the neutral component can decay to $ZZ/Z\gamma/\gamma\gamma$.  
At the 13 TeV LHC, the dominant
production is $pp\rightarrow \Pi^{(1, 3)_0^\pm} \Pi^{(1, 3)_0^0}$ and
has the cross section of around $5$~fb for $M_{\Pi^{(1, 3)_0}}\approx
400$~GeV. After the subsequent decays, we have a collider signature with a pair of electroweak boson resonances with an equal mass at 400 GeV. The LHC Run II should be
able to discover a moderately light electroweak triplet mesons in this model \cite{Freitas:2010ht}.

\subsubsection{Superbumps for a Low Confinement Scale: 750 GeV Diphoton Resonances}
\label{sec:low-scale}
For a low confinement scale, the SM singlet mesons could have strong
enough interactions with two SM gauge bosons via the triangle-anomaly
interactions in Eq.~\eqref{eq:anomaly-meson} to be observed as singly
produced narrow resonances at the LHC.  The scalar and pseudo-scalar
of $\Pi^{1^A}$ have nearly generate masses and similar production
cross sections. As a result, if they are within the collider reach, we
anticipate a superbump signature with two narrow resonances adjacent
to each other. 

Recent hints of a diphoton resonance with a mass of 750 GeV by ATLAS~\cite{ATLAS-CONF-2015-081,ATLASmoriond} and CMS~\cite{CMS-PAS-EXO-15-004,CMSmoriond} could serve as the first superbumps. There is a mild hint from ATLAS that this excess is broader than would be predicted for a resonance with a width less than the detector resolution, though this hint is not aided by a combination with CMS data or Run I data. A superbump can naturally appear with low statistics as a ``broad'' diphoton bump. The broadness of the bump is not due to a single resonance with large width, but is rather due to a superposition of two nearby resonances, each with a narrow width.\footnote{For a similar fit using sgoldstino with a mass splitting, see Refs.~\cite{Petersson:2015mkr,Casas:2015blx,Ding:2016udc,Bardhan:2016rsb}, as well as Ref.~\cite{Baratella:2016daa} for pointing out a potential problem for this interpretation.} 

To fit the observed excess of events near $750~\mbox{GeV}$ at ATLAS and CMS with the signal cross section of around 5 fb and the 45 GeV mass difference of two narrow resonances, we calculate the mass spectrum and mixings using the superpotentials in Eq.~(\ref{eq:ir-superpotential}) and Eq.~(\ref{eq:breaking-super}). We perform a full numerical calculation, avoiding the approximations of the previous sections, other than to assume $\kappa_3 F \ll
\Lambda^2$ and $m_{2,3} \ll \Lambda$, which we will see is a good
approximation for the parameter space of interest. Assuming the non-calculable order-one parameters to be one, there are four parameters for the model: $\kappa_3\,F$, $\Lambda$, $m_2$ and $m_3$.  Three of them can be determined by
the experimental information: the average $\Pi^{1^A}$ meson mass is
$750~\mbox{GeV}$; the mass difference in the $\Pi^{1^A}$ scalar and
pseudo-scalar is $45~\mbox{GeV}$; the sum of the scalar and
pseudo-scalar production cross sections of $\sigma(pp \to \Pi_{R,
  I}^{1^A} \to \gamma\gamma) \approx
5~\mbox{fb}$\cite{Buckley:2016mbr}. We are left with
one more free model parameter, which can be taken to be $m_3$.

\begin{figure}[th!b]
  \centering
\includegraphics[width=0.6\textwidth]{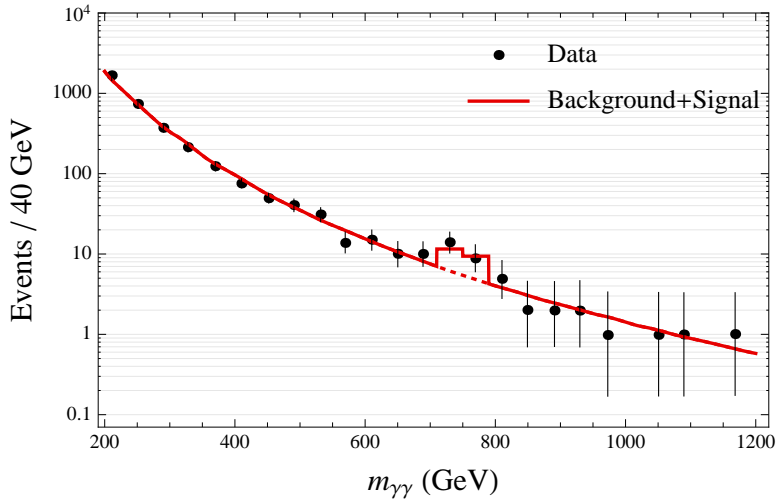}\hspace{0.3cm}
  \caption{A comparison of one benchmark model point with $m_2 = m_3= 850$~GeV, $\kappa_3 \, F = (160~\mbox{GeV})^2$, and $\Lambda=2.5$~TeV and the ATLAS data with 3.2 fb$^{-1}$ luminosity~\cite{ATLAS-CONF-2015-081} at the 13 TeV LHC. The total signal production cross section times diphoton branching ratio is 4.4 fb.
  }
  \label{fig:spectrum}
\end{figure}

Under the narrow resonance approximation, the formula for the
production cross section of singlet mesons is calculated to be
\beqa
\sigma(gg \rightarrow \Pi_{R, I}^{1_A}) = \frac{\pi^2}{8}\, \frac{\Gamma_{\Pi_{R, I}^{1_A} \rightarrow g g }}{M_{\Pi_{R, I}^{1_A}}}\, \delta\left(\hat{s} - M^2_{\Pi_{R, I}^{1_A}}\right)  \,.
\eeqa
We use the MSTW parton distribution function~\cite{Martin:2009iq} to obtain the numerical values of production cross sections at the 13 TeV LHC. In the left panel of Fig.~\ref{fig:spectrum}, we show a comparison of our model with the ATLAS data for a benchmark model point with $m_2 = m_3= 850$~GeV, $\kappa_3 \, F = (160~\mbox{GeV})^2$, and $\Lambda=2.5$~TeV. For this benchmark point, the singlet meson masses are: $M_{\Pi_R^{1^A} } \approx 770$~GeV and $M_{\Pi_I^{1^A} } \approx 730$~GeV. From Eq.~(\ref{eq:singlet-split-MSSM}) and unless the gluino mass in the MSSM is very heavy, the singlet meson mass splitting should mainly come from additional hidden-sector SUSY-breaking effects to have the splitting around 45 GeV. Both $\Pi_R^{1^A}$ and $\Pi_I^{1^A}$ decaying branching ratios should match to the case with $m_2\approx m_3$ in Table~\ref{tab:branching}. 
The summed production cross section times the diphoton branching ratio is around 4.4 fb with 2.0 fb for $\Pi_R^{1^A}$ and 2.4 fb for $\Pi_I^{1^A}$. The ATLAS estimated background is used by us to compare our model predictions with their data at the 3.2 fb$^{-1}$ luminosity~\cite{ATLAS-CONF-2015-081}. For the signal numbers of events, we have multiplied the tree-level results by an additional 70\% signal efficiency~\cite{ATLAS-CONF-2015-081} and ignored the potentially large K-factor because it is within the additional non-perturbative model uncertainties.
For this benchmark model point and before turning on SUSY-breaking effects, we have all meson and mesino masses to be approximately degenerate and around 750 GeV. After the SUSY-breaking effects, the SM-charged meson masses receive additional loop-level corrections. For QCD-singlet mesons, the mass splitting follows the formulas in Table~\ref{tab:splitting}, so the mass splittings for $\Pi_{(1,3)_0}$ and $\Pi_{1_B}$ have the same magnitude as $\Pi^{1^A}$ and are  around 40~GeV and $-40$~GeV, respectively. For the QCD-charged mesons, on the other hand, the MSSM SUSY-breaking contributions dominant and have the mass splitting for the color-octet to be around 400 GeV for $M_{\tilde{g}} = 2$~TeV from Eq.~(\ref{color-split-MSSM}), up to some order-one numbers from non-perturbative physics. We also note that there are additional SUSY-breaking contributions to the overall QCD-charged meson masses, ${\cal O}(\frac{\alpha_s}{\pi} \, M_{\tilde{g}})$, in the K\"ahler potential. The color-octet and triplet mesons are thus a few hundred GeV heavier than the around 750 GeV states, which should be tested at the LHC Run II. 

For the simple {\small 24}MSSM model, in order to fit the 750 GeV diphoton signal with around 5 fb cross-section, the Yukawa coupling is required to be 2.6. For such a large value of the Yukawa coupling, the singlet mass splitting is expected to be several hundred GeV. This splitting is too large to explain the ``broad width'' hint from ATLAS. Furthermore,  such a large coupling runs very quickly to a low scale Landau pole at a scale of a few TeV. This simple model cannot easily to fit the diphoton excess, although it can provide superbump signatures with different masses and production cross sections.  Additional states coupling to the 24 superfield could potentially enhance the cross-section and reduce the required Yukawas \cite{Martin:2009bg}.

\section{Discussion and Conclusions}
\label{sec:conclusion}
The fermion degrees of freedom in our two models have also interesting phenomenological consequence. Concentrating on the SQCD model and depending on the kinematics, the SM-charged mesinos can decay into MSSM gauginos plus their corresponding (most likely off-shell) mesons. The dominant decay channel could be one SM-singlet mesino plus SM gauge bosons via the transition anapole operator in Eq.~(\ref{eq:anapole-charge-radius}). For instance, the color-octet mesino can have the decay of $\widetilde{\Pi}^{(8,1)_0} \rightarrow \widetilde{\Pi}^{1_A} + g$. In the case that the lightest mesino mass is heavier than the LSP, $\widetilde{\chi}^0$, in the MSSM sector. One could have the decay of $\widetilde{\Pi}^{1_A} \rightarrow \widetilde{\chi}^0 + \gamma (Z)$. So, after the QCD-charged mesinos are produced in pair at the LHC, the signature could be $2\,\gamma(Z) + 2\,j + \mbox{MET}$. In the limit of $m_2 = m_3$, the missing transverse energy and $p_T$'s of photons and jets are suppressed because of the nearly degenerate mesino spectrum. In order to discover the superpartners of mesons, one should relax the cuts imposed in MSSM superpartner searches~\cite{Aad:2016jxj}.  

In the other case with the lightest mesino lighter than the LSP in the MSSM, the lightest mesino could be a WIMP candidate. If the electric-neutral component of $\widetilde{\Pi}^{(1,3)_0}$ is the LSP, the situation is very much like the wino case in the MSSM. The main annihilation channel is into two $W$ gauge bosons. For $M_{\widetilde{\Pi}^{(1,3)_0}} \lesssim 1$~TeV, the annihilation cross section is too large to have a thermal dark matter candidate. If the SM-singlet mesino, $\widetilde{\Pi}^{1_A}$, is the LSP. One of the leading annihilation channels is mediated by a scalar or pseudo-scalar in the $s$-channel and has two gluons as the annihilation products. In the limit of $m_2=m_3$, the annihilation cross section scales as $\alpha_s^2  M^{2/3}_{\widetilde{\Pi}^{1_A}}/(16\pi\, \Lambda^{8/3})$, which is too small for providing a thermal WIMP, providing $\Lambda \gtrsim 1$~TeV. One could consider the co-annihilation of $\widetilde{\Pi}^{1_A}$ with other SM-charged mesinos or introduce additional interactions with other MSSM particles to enhance the effective annihilation rate~\cite{Bellazzini:2011et}. 

In our study, we have concentrated on the superbumps with multiple SM gauge bosons as the final-state particles. There are also interesting models with the superbump resonances decaying into two SM fermions. For instance, additional inert Higgs doublets with suppressed VEV's could dominantly decay into two leptons. The LHC may discover such superbumps in multi-lepton final states. The natural mass splitting between the scalar and pseudo-scalar is $\alpha_2\, M_{\widetilde{W}}/(4\pi)\approx 2$~GeV for the $SU(2)_W$ gaugino mass $M_{\widetilde{W}}$ around 1 TeV. If such di-fermion superbumps can be discovered at the LHC, it is straightforward to determine the CP properties of the two states by studying the lepton angular distributions. This would be a clear signal of supersymmetry.

\subsection*{Acknowledgments}
We would like to thank Vernon Barger and Lisa Everett for discussion and comments, as well as Ran Lu for the suggestion of the title of this paper. This work is supported by the U. S. Department of Energy under the contract DE-FG-02-95ER40896.

\begin{appendix}
\section{Anomaly-induced Superpartner Couplings}
\label{appendix:other-couplings}
In this Appendix, we determine the component couplings determined by
Eq.~\eqref{eq:loop-interaction}.  We begin by expanding $\int d^2\theta\,
\mathbf{\Pi} \,\mathbf{W}^{a\,\alpha}_V \, \mathbf{W}^a_{V,\alpha} + {\rm h.c.}$ in
components, assuming that it has a real coefficient.  We write the
components generally as
\beqa
\mathbf{\Pi} & = & \frac{\Pi_R + i\Pi_I}{\sqrt{2}} + \sqrt{2}\,
  \theta \,\widetilde{\Pi} + \theta^2 F_\Pi  \,,\nonumber\\
  \mathbf{W}^a_{V,\alpha} & = & \lambda^a_{\alpha} + \theta_\alpha D^a_V +
  \frac{i}{2} (\sigma^\mu \bar{\sigma}^\nu \theta)_\alpha V_{\mu\nu}^a
  + i \,\theta\theta (\sigma^\mu \partial_\mu \lambda^{a\dagger})_\alpha\,. 
\eeqa
where $\widetilde{\Pi}$ is the mesino and $\lambda$ is the
gaugino corresponding to $V$.  For convenience, we also define $\tilde{\Psi}^a = (\tilde{\psi}_\alpha^a,
\tilde{\psi}^{a\dagger\dot{\alpha}})$, the four-component notation for
fermions using upper-case rather than lower-case letters.  Then, in
components, we find
\beqa
\int d^2\theta\, \mathbf{\Pi} \,\mathbf{W}^{a\,\alpha}_V\,
\mathbf{W}^a_{V,\alpha} + {\rm h.c.} & = & \sqrt{2}\, i \Pi_R
\overline{\lambda}^a \slashed{\partial} \lambda^a + \sqrt{2} \,\Pi_I
\overline{\lambda}^a \gamma^5 \slashed{\partial} \lambda^a -
\frac{\Pi_R}{\sqrt{2}} \,V^{a\mu\nu} V^a_{\mu\nu} -
\frac{\Pi_I}{\sqrt{2}}\, V^{a\mu\nu} \widetilde{V}^a_{\mu\nu} \nonumber\\
& + & \sqrt{2}\,
\Pi_R (D_V^a)^2 - \frac{i}{\sqrt{2}} \,\overline{\widetilde{\Pi}} \gamma^\mu
\gamma^\nu \lambda^a V^a_{\mu\nu} - \sqrt{2}\, \overline{\widetilde{\Pi}}
\lambda^a D^a \nonumber\\
& + & F_\Pi \,\overline{\lambda}^a P_L \lambda^a +
F_\Pi^\dagger\, \overline{\lambda}^a P_R \lambda^a \,.
\eeqa
The general form of the couplings of mass basis fields can be obtained
by then using $\mathbf{\Pi} = \mathbf{H}^\prime, \mathbf{\Pi}^{24}$
and $V = B, W, G$, the coefficients from Tables~\ref{tab:coefficients-hp}, and the
mixing angles in Eq.~\eqref{eq:mixing-matrix}.  SUSY breaking effects are neglected in these
expressions.  They will generally be model dependent as discussed in
the main text.  As an example of this procedure for obtaining the
general form of the couplings, we present the full form of the scalar
$\Pi^{1^A}$ coupling to gluons as
\beqa
&&\frac{1}{16 \cdot 2^{7/8} \cdot 3^{1/8}  \sqrt{5} \pi
  \sqrt{m_2} (\Lambda m_3)^{3/4}} \left[ \left(3 (m_2 +
    2m_3) \cos\theta - \sqrt{6}\, (m_3 - 2 m_2) \sin\theta\right)
\right. \nonumber\\
& & \left.  \hspace{0.5cm} + \, \left(2 (m_2 +
    2 m_3) \cos\theta - \sqrt{2/3} \,(4 m_3 - 3 m_2) \sin\theta\right) 
    +\,
    \left(m_3
    \cos\theta - \sqrt{2/3} \,m_3 \sin\theta\right)  \right] \,.
\eeqa

The coupling of mesons to gauginos is given by
\beqa
{\cal L} \supset - c_A \frac{\sin\theta \sqrt{N_f}}{\pi \,
  \Lambda}\, \left[ \Pi^{1^A}_R \sum_i 2 \,g_i^2\, i \overline{\lambda_i^a}
    \slashed{\partial} \lambda_i^a  - \Pi^{1^A}_I \sum_i 2\, g_i^2\, \overline{\lambda_i^a}
    \slashed{\partial} \gamma^5 \lambda_i^a \right].
\eeqa
The coupling of the mesons to the $D$-term is given by
\beq
{\cal L} \supset - c_A \frac{\sin\theta\sqrt{N_f}}{\pi\,\Lambda}\,
\Pi^{1^A}_R \sum_i 2 g_i^2\, D_i^a\, D_i^a\,.
\eeq
While the first term induces couplings only to squarks, the second and
third terms can induce couplings to the Higgs fields. In the decoupling limit of heavy Higgs fields and with $\alpha \approx \beta - \pi/2$, the Higgs contribution to the D-term is given by~\cite{Martin:1997ns}
\beq
 g_2^2 \, D_2^i \, D_2^i + g_Y^2 \, D_Y^2 = \frac{1}{16}\,
 (g_2^2 + g_Y^2)\, (h+\sqrt{2}\,v)^4 \, \cos^2 2\beta \,,
\eeq
with $v\approx 174$~GeV as the electroweak VEV. This coupling induces a tadpole for the scalar, as well as a mass mixing with the Higgs boson, which are both negligible for phenomenological consequence. The coupling of the fermionic partner to the meson, which we call the mesino, to gauge bosons and gauginos
is given by 
\beqa
{\cal L} \supset - c_A \frac{\sin\theta \sqrt{N_f}}{\pi \,
  \Lambda} \sum_i 2 i
 \,  g_i^2 \, \overline{\lambda_i^a} \, \gamma^\mu \, \gamma^\nu \,
 \gamma^5\, \widetilde{\Pi}^{1^A} \,  F^a_{i\,\mu\,\nu} \,.
\eeqa
The coupling of the mesino to the $D$-term is given by
\beq
{\cal L} \supset - c_A \frac{\sin\theta \sqrt{N_f}}{\pi \, \Lambda}
\sum_i -2 g_i^2\, \overline{\widetilde{\Pi}}^{1^B} \lambda_i^a D_i^a\,.
\eeq
Finally, the coupling of the meson $F$-term is given by
\beq
{\cal L} \supset - c_A \frac{\sin\theta \sqrt{N_f}}{\pi \, \Lambda}
\sum_i \frac{1}{\sqrt{2}} g_i^2 [F \overline{\lambda_i^a} P_L
\lambda_i^a + F^* \overline{\lambda_i^a} P_R \lambda_i^a] \,.
\eeq

\section{Supersymmetric K{\"a}hler Potential}
\label{sec:kahler-susy}
For the K{\"a}hler potential, there is no non-renormalization theorem to constrain the superfield interactions. So, all operators consistent with the symmetries in our model should appear. One may worry about the additional un-controllable modifications to our model spectrum. In this section, we want to show that this is not the case and there is additional suppression factors in powers of $m_{2,3}/\Lambda$. At dimension-four level, we have the following non-derivative interactions of the mesons in the effective K{\"a}hler potential in the low energy theorem~\cite{Cho:1998vc}
\beqa
K_{\rm eff} = - \frac{(4\pi)^2}{\Lambda^2}\,\left[ \frac{\kappa_1}{N_f}\,\mbox{Tr}({\bf M}^\dagger {\bf M}\, {\bf M}^\dagger {\bf M} ) \, + \, \frac{\kappa_2}{N_f^2}\,\mbox{Tr}({\bf M}^\dagger {\bf M}) \mbox{Tr}({\bf M}^\dagger {\bf M} )
\right] \, + \, \cdots \,. 
\eeqa
Here, $\kappa_{1,2}$ are numbers order of unity. Here, we don't include operators containing the baryon superfields, which will change the meson superfield masses. Taking the $\theta\theta\theta^\dagger \theta^\dagger$ component and keeping the non-derivative terms, we have 
\beqa
V_{K} &=& \frac{(4\pi)^2}{\Lambda^2} \left[
\frac{\kappa_1}{N_f}\left(
2\,\mbox{Tr}[{\rm M}^\dagger {\rm M} F_{\rm M}^\dagger F_{\rm M} ] \,+\, 
2\,\mbox{Tr}[{\rm M} {\rm M}^\dagger  F_{\rm M} F_{\rm M}^\dagger ] 
 \right) \right.
\nonumber \\
&&\left.  \hspace{1.0cm} \,+\,
\frac{\kappa_2}{N_f^2}\left(
2\,\mbox{Tr}[{\rm M}^\dagger {\rm M} ]\mbox{Tr}[F_{\rm M}^\dagger F_{\rm M} ] 
\,+\, 2\,\mbox{Tr}[{\rm M}^\dagger F_{\rm M} ]\mbox{Tr}[F_{\rm M}^\dagger {\rm M} ]  \right)
\right] \,.
\eeqa
In the limit of $m_{2}=m_{3} \ll \Lambda$, it is easy to show that the dominant contributions to meson masses goes as
\beqa
\frac{\Delta M_{\Pi_i}^K}{M_{\Pi_i}} = {\cal O}\, \left[ \left( \frac{m_3}{\Lambda} \right)^{1/2} \right] \,.
\eeqa
For the baryons, a similar estimation shows that the correction from the K{\"a}hler potential is suppressed by $(m_3/\Lambda)^{3/2}$. 

\section{SUSY-breaking K\"ahler Corrections}
\label{sec:kahler-susy-breaking}
In additional to the superpotential soft mass discussed at length in
Section \ref{sec:susy-breaking}, there are also K\"ahler potential soft masses
that have the form
\beq
{\cal L} \supset \int d^4\theta \frac{\mathbf{X}^\dagger
  \mathbf{X}}{M_{\rm messenger}^2} \left(\kappa_i \mathbf{Q}^\dagger_i 
  \mathbf{Q}_i + \tilde{\kappa}_i \mathbf{\widetilde{Q}}^\dagger_i
  \mathbf{\widetilde{Q}}_i \right),
\eeq
where $\kappa_i$ and $\tilde{\kappa}_i$, $i = 2,3$, are real coefficients.
In the low energy theory and to leading order in the superfields, such terms
generate K\"ahler terms of the form
\beq
{\cal L} \supset \int d^4 \theta \frac{\mathbf{X}^\dagger
  \mathbf{X}}{M_{\rm messenger}^2} \left(\kappa_{M,ij}\, \mathbf{M}^\dagger_{ij} 
  \mathbf{M}_{ij} + \kappa_{B,i}\, \mathbf{B}_i^\dagger \mathbf{B} + \kappa_{\tilde{B},i} \,\mathbf{\widetilde{B}}^\dagger_i
  \mathbf{\widetilde{B}}_i\right).
\eeq
Soft mass terms of this form do not split the scalar and pseudo-scalar
meson masses since they couple as $\phi^* \phi = (\phi_R^2 + \phi_I^2)/2$.
Nevertheless they can give corrections to the scalar spectrum that
were not included in the analysis above.  These coefficients can also
be negative, leading to a scalar spectrum that is lighter than the
fermion spectrum.  The corrections to the potential should be much
smaller than the confinement scale such that the SQCD analysis above
remains valid.  In other words, we need
\beq
\kappa_I \frac{F}{M_{\rm messenger}} \ll \Lambda  \,,
\eeq
for $\kappa_I = \kappa_{M,ij},\,\kappa_{B,i},\,\kappa_{\tilde{B},i}$.

\end{appendix}

\bibliography{SUSYDisc}
\bibliographystyle{JHEP}
 \end{document}